\documentclass[final,3p,times]{elsarticle}

\usepackage{amssymb}
\usepackage{amsmath}

\usepackage{graphicx}\usepackage{multirow}\usepackage{amsmath,amssymb,amsfonts}\usepackage{amsthm}\usepackage{mathrsfs}\usepackage[title]{appendix}\usepackage{xcolor}\usepackage{textcomp}\usepackage{manyfoot}\usepackage{booktabs}\usepackage{algorithm}\usepackage{algorithmicx}\usepackage{algpseudocode}\usepackage{listings}\usepackage{newfloat}
\usepackage{soul}
\usepackage{subcaption}
\usepackage{comment}
\usepackage{enumitem}
\usepackage{tikz}
\usepackage{makecell}
\usepackage{url}
\usepackage{adjustbox}
\usepackage[graphicx]{realboxes}
\usepackage{hyperref}

\newcommand\sbullet[1][.5]{\mathbin{\vcenter{\hbox{\scalebox{#1}{$\bullet$}}}}}

\newcommand{\multicomment}[1]{}

\newcommand{\subr}[1]{{\small\texttt{r/#1}}}
\newcommand{\subrtab}[1]{{\footnotesize\texttt{r/#1}}}

\newcommand{\rev}[1]{\textcolor{black}{#1}}

\begin{document}

\begin{frontmatter}

\title{Investigating the heterogeneous effects of a massive content moderation intervention via Difference-in-Differences}

\author[1,3]{Lorenzo Cima}
\author[2,3]{Benedetta Tessa}
\author[3]{Amaury Trujillo}
\author[3]{Stefano Cresci}
\author[1]{Marco Avvenuti}

\affiliation[1]{organization={Dept. of Information Engineering, University of Pisa},addressline={Largo Lazzarino 1}, 
            postcode={56122}, 
            city={Pisa},
            country={Italy}}
\affiliation[2]{organization={Dept. of Computer Science, University of Pisa},addressline={Largo Bruno Pontecorvo 3},
            postcode={56127}, 
            city={Pisa},
            country={Italy}}
\affiliation[3]{organization={IIT-CNR},addressline={Via Moruzzi 1}, 
            postcode={56124}, 
            city={Pisa},
            country={Italy}}

\begin{abstract}
In today's online environments, users encounter harm and abuse on a daily basis. Therefore, content moderation is crucial to ensure their safety and well-being. However, the effectiveness of many moderation interventions is still uncertain. Here, we apply a causal inference approach to shed light on the effectiveness of The Great Ban, a massive social media deplatforming intervention on Reddit. We analyze 53M comments shared by nearly 34K users, providing in-depth results on both the intended and unintended consequences of the ban. Our causal analyses reveal that 15.6\% of the moderated users abandoned the platform while the remaining ones decreased their overall toxicity by 4.1\%. \rev{Nonetheless, a small subset of users exhibited marked increases in both the intensity and volume of toxic behavior, particularly among those whose activity levels changed after the intervention. However, these reactions were not accompanied by greater activity or engagement, suggesting that even the most toxic users maintained a limited overall impact.} Our findings bring to light new insights on the effectiveness of deplatforming moderation interventions. Furthermore, they also contribute to informing future content moderation strategies and regulations.
\\\\ \textcolor{red}{Article published in \textit{Online Social Networks and Media (OSNEM)}. DOI: \href{https://doi.org/10.1016/j.osnem.2025.100320}{10.1016/j.osnem.2025.100320}. Please, cite the published version.}
\end{abstract}


\end{frontmatter}

\section{Introduction}
\label{sec:introduction}

Content moderation is a fundamental component of many online platforms as it prevents the spread of problematic content and behavior, such as hate speech~\citep{giorgi2024human,kocon2021offensive,yin2022hidden}, fake news ~\citep{bazmi2023multi,augenstein2024factuality,corsi2024crowdsourcing}, and polarization ~\citep{tao2025detecting,trujillo2023one,donkers2023sounding}.
Content moderation also contributes to the fairness and safety of online users by promoting high ethical standards and fostering the health of online communities~\citep{gillespie2018custodians}. In practice, administrators enforce platform policies through the application of a wide array of moderation interventions, which can vary in severity~\citep{trujillo2023dsa}.
For example, moderators can send short warning messages and use informative labels~\citep{katsaros2022reconsidering} or even take the drastic decision to remove content and/or users~\citep{trujillo2022make,jhaver2021evaluating,russo2023spillover}.
However, despite the increasing reliance on content moderation, there is still little understanding of the effects of many interventions, which hinders our capacity to design and apply effective moderation.
For example, recent studies demonstrated that while some interventions had the desired effects, others resulted in heterogeneous~\citep{horta2021platform,trujillo2023one} or even ineffectual outcomes~\citep{dias2020emphasizing}, or even led to undesirable consequences~\citep{bail2018exposure,pennycook2020implied}.
Therefore, assessing the effects of moderation interventions represents an important preliminary step towards developing new and more effective forms of moderation.

The most popular and widely used type of intervention is \textit{deplatforming}, which involves the removal of content, users, or even entire communities~\citep{ribeiro2024deplatforming}.
Notorious examples are the ban that Donald Trump received in 2021
from Facebook and X (formerly Twitter)~\citep{seeliger2023twitter} and the deplatforming of three particularly toxic influencers from X ~\citep{jhaver2021evaluating}. 
Additionally, X removed accounts involved in coordinated inauthentic behavior~\citep{cima2024coordinated} and Reddit permanently shut down different communities because of racism, sexism, and hatefulness~\citep{chandrasekharan2017you,habib2022proactive}.
In June 2020, Reddit itself hosted one of the biggest deplatforming campaigns in the history of social media --\textit{The Great Ban}-- which resulted in around 2,000 subreddits being banned due to ongoing spread of toxicity and hate speech.\footnote{\url{https://www.reddit.com/r/announcements/comments/hi3oht/update_to_our_content_policy/} (accessed: 10/15/2024)}
Among these are popular communities such as \subr{The\_Donald} and \subr{ChapoTrapHouse}.  
Despite its impact on several communities and users within and outside Reddit, its effects are still little-explored. 
For example, current research on The Great Ban has mainly investigated the changes in the writing style of the users~\citep{trujillo2021echo} without assessing the ban's effectiveness. Few studies analyzed the changes in toxicity, and those who did focused only on a small set of subreddits~\citep{trujillo2022make}. Moreover, the majority of the existing works on The Great Ban are focused on community-level effects, neglecting the individual user-level responses that are instrumental for understanding how effective the ban was in mitigating problematic behaviors~\citep{robertson2022uncommon,trujillo2023one}. In a previous work, we provided preliminary results on the effectiveness and unintended consequences of The Great Ban~\citep{cima2024great}. Here, we extend our previous analysis by adopting a robust causal inference approach to estimate the effects of the intervention. Specifically, we employ a Difference-in-Differences (DiD) methodology, which allows us to quantify changes in user behavior while accounting for general platform-wide trends. To do so, we construct a large-scale dataset comprising 53 million comments from nearly 34,000 users, covering the seven months before and after the ban. We define a focus group of users who were active in the banned subreddits and compare their post-ban behavior to that of a baseline group of thematically related but unaffected subreddits. We then analyze changes in toxicity, activity, and engagement to assess both the intended and unintended consequences of the Great Ban. With respect to our previous work we also address an additional research question investigating the impact that toxic users had after the intervention, as explained in the following.

\textbf{Research focus.} We address the outstanding knowledge gaps by conducting a comprehensive quantitative causal analysis of the changes in toxicity among users active in the 15 most popular subreddits involved in The Great Ban. We analyze 53M comments posted by nearly 34k users over a period of 14 months guided by the following research questions.

$\sbullet[.75]\;$\textbf{RQ1:} \textit{Did The Great Ban effectively reduce toxicity?}
Studies have shown that some interventions led to an increase rather than a decrease in toxic behavior. Here we assess the effectiveness of The Great Ban in reducing toxicity, as hate and toxic speech were the main reasons behind the ban.

$\sbullet[.75]\;$\textbf{RQ2:} \textit{Did The Great Ban lead to any unintended side effects for certain users?} That is, \textit{were there users who became significantly more toxic after the intervention?}
The evaluation of the outcomes of a moderation intervention has to take into consideration the possible presence of users who grew resentful of the platforms and increased -- rather than decreased -- their toxicity. Such extreme reactions can arise even amid an overall reduction in toxicity at platform- or community- level, requiring further analyses at \textit{user level}. In this study, we evaluate and estimate the extent of these reactions to The Great Ban, considering them as potential side effects of the intervention.

$\sbullet[.75]\;$\textbf{RQ3:} \textit{Did the banned subreddits foster toxicity by driving engagement and activity?} To deepen our analysis, we examine whether the banned subreddits played a role in amplifying toxic behaviors by analyzing the activity levels and engagement received by users who became significantly more toxic after the ban. Understanding activity is crucial, as highly toxic users with low activity may have a limited overall impact, whereas those who are both highly active and highly toxic can pose a greater risk to platform integrity. Additionally, analyzing engagement allows us to assess whether toxic behaviors were discouraged or reinforced within certain parts of Reddit. If toxic users consistently attracted attention and interactions, it may suggest that their behavior was not only tolerated but actively incentivized by other users~\citep{jiang2023social}.

\textbf{Main findings.} Based on the insights gained from answering the previous research questions, our study produces the following main findings:
\begin{itemize}
    \item The Great Ban led 15.6\%  of the moderated users to abandon the platform, while those who stayed reduced their toxicity by an average of 4.1\%.
    \item While the reduction in toxicity was limited, a significant fraction of users became much more toxic. In particular, 5\% of users increased their toxicity by over 70\% compared to their pre-ban levels.
    \item Resentful users who escalated their toxicity were found across each of the analyzed subreddits. However, their impact was limited probably due to their limited activity and the lack of positive feedback from other users.
\item Major changes in a dimension of user behavior do not necessarily lead to major changes in other dimensions. For example, the majority of users who drastically increased their toxicity did not experience significant changes in activity or engagement.

\end{itemize}
Our work provides an in-depth analysis of the effects of The Great Ban. It points out its shortcomings by highlighting the complex challenges of moderating different communities. As such, our findings can inform the development of future moderation interventions.
 
\section{Related Work}
\label{sec:related-work}
We review and analyze recent literature on the evaluation of moderation interventions, beginning with studies most closely related to our own.
\subsection{Deplatforming}
\label{sub:deplatforming}
The Great Ban was a significant event that had a major impact on Reddit.
However, only a few studies actually looked into its effects.
One of them is the study conducted by M.~\citet{trujillo2021echo}, that investigated the changes in activity and use of language among the 15 most popular subreddits involved in the ban.
They discovered that the most active users drastically decreased their level of activity and observed heterogeneous responses to the ban both intra- and inter-subreddits.
Here, we build on this study by evaluating the impact of The Great Ban in terms of toxicity, rather than focusing on activity and language. Another body of works focused on assessing the effects of deplatforming in a subset of subreddits affected by the ban or in entirely different subreddits.
\citet{chandrasekharan2017you} and \citet{saleem2018aftermath} evaluated how The Great Ban impacted two specific subreddits: \subr{fatpeoplehate} and \subr{coontown}. They revealed how a large portion of users abandoned the platform and the remaining ones notably decreased their toxicity levels.
However, they also observed that several users migrated to other subreddits and doubled their posting activity~\citep {chandrasekharan2022quarantined}.
Other studies focused on deplatforming both within and outside Reddit.
\citet{horta2021platform} studied the migration of Reddit users from banned subreddits to other platforms after a deplatforming intervention. They show how migrated users became drastically less active on the new platforms. This was not the case for a group of users, who instead became much more toxic and radicalized. 
~\citet{mekacher2023systemic}
analyzed the consequences of deplatforming from Twitter to Gettr. They found that politically polarized users exhibit lower toxicity on fringe platforms, likely due to reduced exposure to interactions with out-group members. Moreover,~\citet{jhaver2021evaluating} studied the impact of the banishment of three notorious influencers from Twitter. They observed a general decrease in attention towards these influencers and toxicity among their fans. However, some users became much more active and toxic. Finally,~\citet{cima2024coordinated,schoch2022coordination}, and~\citet{seckin2024labeled} examined large-scale bans implemented by Twitter to combat coordinated inauthentic behaviors. While these studies did not assess the effects of such bans, they tackled the challenging task of constructing a control group to compare the behavior of banned users (i.e., the treatment group). Their findings underscore the inherent difficulties in identifying suitable control groups for causal analyses in online moderation studies.

In general, the previous studies highlight how moderation interventions do not always yield the desired results and that different communities and different users can react differently to the same intervention. Our work aims to extend the existing literature by shedding light on the effects of The Great Ban, one of the biggest -- yet little-explored -- moderation interventions. Our study provides a comprehensive evaluation of the ban's impact on the 15 most popular subreddits in terms of toxicity, activity, and engagement.

\begin{table*}[t]
    \footnotesize
    \setlength{\tabcolsep}{2pt}
    \centering
    \adjustbox{max width=\textwidth}{
    \begin{tabular}{lrcrrcrrcrr}
	\toprule
	&&& \multicolumn{2}{c}{\texttt{IN-BEFORE}} && \multicolumn{2}{c}{\texttt{OUT-BEFORE}} && \multicolumn{2}{c}{\texttt{OUT-AFTER}} \\
	\cmidrule(lr){4-5} \cmidrule(lr){7-8} \cmidrule(lr){10-11}
        \textbf{subreddit} & \textbf{subscribers} && \textit{core users} & \textit{comments} && \textit{users} & \textit{comments} && \textit{users (\%)} & \textit{comments (\%)} \\
        \midrule
        \subrtab{chapotraphouse} & 159,185 && 9,295 & 1,368,874 && 9,205 & 3,947,894 && 8,319 (90.37) & 3,157,462 (79.98) \\
        \subrtab{the\_donald} & 792,050 && 4,262 & 619,434 && 4,132 & 2,578,026 && 3,145 (76.11) & 1,434,008 (55.62) \\
        \subrtab{darkhumorandmemes} & 421,506 && 1,632 & 35,561 && 1,617 & 1,246,399 && 1,392 (86.09) & 689,079 (55.29) \\
        \subrtab{consumeproduct} & 64,937 && 1,730 & 60,073 && 1,719 & 1,209,933 && 1,275 (74.17) & 594,349 (49.12) \\
        \subrtab{gendercritical} & 64,772 && 1,091 & 94,735 && 1,039 & 511,173 && 706 (67.95) & 287,877 (56.32) \\
        \subrtab{thenewright} & 41,230 && 729 & 5,792 && 726 & 600,057 && 575 (79.20) & 308,584 (51.43) \\
        \subrtab{soyboys} & 17,578 && 596 & 5,102 && 594 & 454,659 && 432 (72.73) & 190,570 (41.91) \\
        \subrtab{shitneoconssay} & 8,701 && 559 & 9,178 && 555 & 338,218 && 384 (69.19) & 140,619 (41.58) \\
        \subrtab{debatealtright} & 7,381 && 488 & 27,814 && 476 & 274,600 && 328 (68.91) & 117,281 (42.71) \\
        \subrtab{darkjokecentral} & 185,399 && 316 & 3,214 && 308 & 307,876 && 270 (87.66) & 179,067 (58.16) \\
        \subrtab{wojak} & 26,816 && 244 & 1,666 && 240 & 210,249 && 170 (70.83) & 81,142 (38.59) \\
        \subrtab{hatecrimehoaxes} & 20,111 && 189 & 775 && 188 & 185,379 && 143 (76.06) & 96,457 (52.03) \\
        \subrtab{ccj2} & 11,834 && 150 & 9,785 && 145 & 101,165 && 119 (82.07) & 63,393 (62.66) \\
        \subrtab{imgoingtohellforthis2} & 47,363 && 93 & 376 && 92 & 74,664 && 72 (78.26) & 43,018 (57.62) \\
        \subrtab{oandaexclusiveforum} & 2,389 && 60 & 1,313 && 59 & 48,774 && 55 (93.22) & 35,853 (73.51) \\
        \midrule
        focus group total (unique) &&& 16,828 & 2,243,692 && 16,540 & 8,235,086 && 13,963 (84.42) & 5,592,321 (67.91) \\
        \midrule
        \subrtab{askreddit} & 44,279,043 && 15,863 & 3,220,750 && 15,850 & 17,014,057 && 15,576 (98.27) & 12,372,977 (72.72) \\
        \subrtab{worldnews} & 34,091,866 && 10,051 & 577,777 && 10,050 & 12,728,027 && 9,907 (98.58) & 9,517,097 (74.77) \\
        \subrtab{politics} & 8,438,064 && 9,002 & 1,862,399 && 8,998 & 11,059,478 && 8,862 (98.49) & 8,328,241 (75.30) \\
        \midrule
        baseline group total (unique) &&& 16,955 & 5,660,926 && 16,937 & 18,480,693 && 16,633 (98.21) & 13,378,316 (72.39) \\
	\bottomrule
    \end{tabular}}
    \caption{Composition of the dataset. Subreddits are sorted by number of \textit{core users} before the ban. Data in \texttt{IN-BEFORE} refers to user activity within the specified subreddits before the ban occurred. \texttt{OUT-BEFORE} and \texttt{OUT-AFTER} represent user activity outside of the specified subreddits (i.e., in the rest of Reddit), respectively before and after the ban. The percentages in the \texttt{OUT-AFTER} columns indicate the fraction of users and comments after the ban with respect to those in \texttt{OUT-BEFORE}.}
\label{tab:dataset}
\end{table*}
 
\subsection{Soft moderation}
\label{sub:light}
Despite deplatforming being one of the most widely adopted moderation interventions, it can be perceived by many users as a threat to free speech, as it involves the removal of content and users~\citep{zannettou2021won}. This concern has led to the rise of an alternative category of interventions known as \textit{soft} interventions, which attempt to mitigate harmful behavior without outright banning users or communities. Recently, these strategies have been the focus of several studies. For example,~\citet{trujillo2022make,trujillo2023one} investigated the effects of quarantines and restrictions, which are moderation measures frequently implemented before community bans on Reddit. Quarantines limit a subreddit’s visibility by removing it from search results and requiring users to opt in before viewing its content, while restrictions can include measures such as disabling specific posting privileges or preventing the subreddit from generating revenue. These interventions serve as intermediate steps to curb harmful behavior before resorting to a full ban. In particular, they analyzed the effects of moderation on \subr{the\_donald}, revealing a decrease in overall activity and toxicity. However, this came at the expense of increased political polarization and decreased factual accuracy in shared news \citep{trujillo2022make,trujillo2023one}. \citet{chandrasekharan2022quarantined} and~\citet{shen2022tale} also studied the quarantine of \subr{the\_donald}, finding that this intervention was largely ineffectual and did not produce notable changes in misogynistic and racist comments or user engagement.

Another common example is flagging disputed posts with warning labels, whose effectiveness has been studied in terms of perceived credibility and user engagement. \citet{pennycook2020implied} found that warning labels can increase the perceived credibility of unlabeled posts, potentially amplifying misinformation that has not been flagged. In addition,~\citet{zannettou2021won} demonstrated that tweets with warning labels often receive higher engagement, as users attempt to debunk the claims. This paradoxical effect suggests that soft moderation can sometimes lead to unintended amplification, rather than suppression, of harmful content. Similarly,~\citet{katsaros2022reconsidering} assessed the effectiveness of preemptive warning messages shown to users before they post toxic tweets. Their A/B test on Twitter confirmed that this strategy was generally effective at reducing toxic content, although a small fraction of users actually increased their toxicity in response to the warnings.

While soft moderation offers a less disruptive alternative to deplatforming, these studies highlight its mixed effectiveness. Unlike hard moderation, soft interventions allow communities to adapt to restrictions rather than being eliminated outright. However, they also risk reinforcing echo chambers by isolating communities without fully dismantling them~\citep{horta2021platform}. For instance,~\citet{chandrasekharan2022quarantined} found that quarantined subreddits often develop stronger internal cohesion and more extreme discourse, which can counteract the intended moderation effects. The trade-offs between soft and hard moderation suggest that no single approach is universally effective~\cite{cresci2022personalized}. Deplatforming is more decisive, removing toxic actors from a platform but potentially leading to migration effects \citep{horta2023deplatforming}. Soft moderation, on the other hand, allows for gradual adaptation, but it can be circumvented, may have limited effectiveness, and can sometimes inadvertently strengthen harmful communities \citep{chandrasekharan2022quarantined}. These nuances highlight the need for hybrid moderation strategies, where soft interventions serve as preliminary measures before escalating to deplatforming when necessary~\cite{kiesler2012regulating}.

Overall, these studies emphasize that both soft and hard moderation interventions come with inherent trade-offs, and additional research is needed to assess how different strategies interact and whether a combination of approaches can yield better moderation outcomes.
 \begin{figure}[t]
        \centering
        \includegraphics[width=0.7\columnwidth]{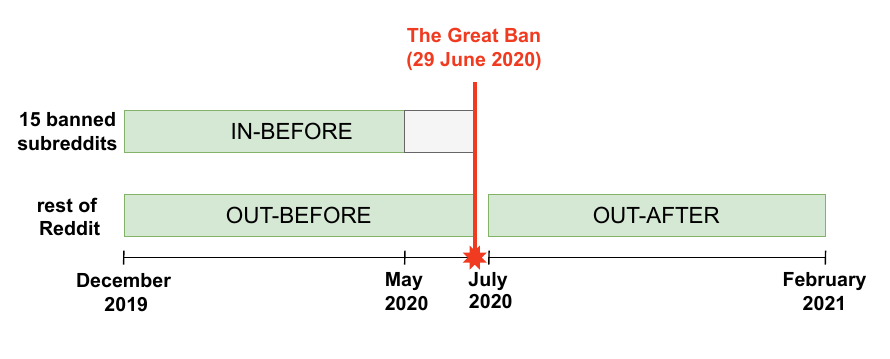}\caption{Timeline of the data collection and analysis periods. Our study covers two 7-month intervals centered around The Great Ban. The \texttt{IN-BEFORE} dataset contains no data after May 2020, suggesting that activity in the banned subreddits ceased prior to the official date of the ban.}
        \label{fig:period}\end{figure}

\section{Dataset}
\label{sec:dataset}
Our dataset\footnote{\url{https://doi.org/10.5281/zenodo.14034510}} includes 16M comments shared by 16,828 distinct Reddit users who have been active in at least one of the 15 most popular public subreddits that were banned during The Great Ban~\citep{trujillo2021echo}, as detailed in Table~\ref{tab:dataset}. In this context, the term \textit{popular} refers to the number of daily active users.
Even though Reddit administrators initially published an obfuscated list of the most popular banned subreddits,\footnote{\url{https://www.redditstatic.com/banned-subreddits-june-2020.txt} (accessed: 10/15/2024)} M.~\citet{trujillo2021echo} were able to decipher this list for subreddits with over 2,000 daily active users. A total of 15 subreddits were identified. The composition of our dataset is shown in Figure~\ref{fig:period}, and the procedure adopted to build it is described below.

\subsection{Focus group} 
We began by collecting all comments posted between December 2019 and June 2020 in each of the 15 popular banned subreddits, yielding 8M comments from 194K distinct users. To this end, we used the Pushshift data archive~\citep{baumgartner2020pushshift}. 
The dataset covers 30 weeks (7 months) leading up to The Great Ban, providing a suitable baseline for user activity before the moderation intervention~\citep{trujillo2023one}. Notably, we were unable to collect data between May and June 2020 due to some subreddits halting their activity or being banned prior to Reddit's public announcement of The Great Ban on June 29, 2020, as shown in Figure~\ref{fig:period}.

Following recent literature~\citep{trujillo2022make, bouleimen2023dynamics},  we identified a representative set of users for the selected subreddits by focusing on \textit{core users}, that is users who consistently participated in at least one subreddit. Core users were characterized as those who posted at least one comment each month from December 2019 to March 2020.\footnote{We excluded April 2020 from this analysis due to the limited number of collected comments, likely because subreddit activity had already slowed or halted in the early days of that month.}
Moreover, we excluded bots (i.e., clearly automated accounts) by removing all accounts that posted two or more comments at the exact same timestamp~\citep{hurtado2019bot}.
To this end, we implemented a minimum time interval of 1 second between comments, ensuring that we did not unintentionally exclude genuine users. This process led to the identification of 741 automated accounts among the 17,569 core users (4.22\%), which we discarded. We verified the validity of this approach by manually validating a random sample of 1,000 comments. The automated accounts exhibited repetitive posting patterns and often engaged in behaviors characteristic of information dissemination bots, such as mass-posting external links or reposting the same messages across multiple threads. After these filtering steps, we ended up with 2.2M comments made by 16,828 core users. Henceforth, we denote this subset of our dataset as \texttt{IN-BEFORE}, representing activity \textit{within} the banned subreddits \textit{prior to} the ban.

To ensure a fair evaluation of the effects of The Great Ban, we need to match comparable datasets before and after the intervention.
Since no activity exists inside the banned subreddits following the intervention, we collected all comments made by the core users outside the 15 banned subreddits during the 7 months before and after the ban as shown in Figure~\ref{fig:period}. All user-level analyses are based on immutable user IDs, ensuring that user activity is accurately tracked across the pre- and post-ban periods. We gathered approximately 13.8M comments from 16,540 distinct users. Data related to user activities before the ban was labeled as \texttt{OUT-BEFORE}, and data after the ban as \texttt{OUT-AFTER}, as shown in Table~\ref{tab:dataset}. Estimates of the ban's effects were derived by comparing the \texttt{OUT-BEFORE} and \texttt{OUT-AFTER} datasets.

\begin{table*}[t]
    \small
\centering
    \begin{tabular}{cclcl}
	\toprule
        \textbf{cluster} && \textbf{banned subreddits} && \textbf{non-banned subreddits} \\
        \midrule
        \multirow{4}{*}{\textit{cluster 1}} && \subrtab{ccj2} && \subrtab{politics} \\      
        && \subrtab{hatecrimehoaxes} && \subrtab{worldnews} \\
        && \subrtab{shitneoconssay} && \subrtab{askreddit} \\
        && \subrtab{soyboys} && \\
        \midrule
        \multirow{3}{*}{\textit{cluster 2}} && \subrtab{debatealtright} && \subrtab{askreddit} \\
        && \subrtab{gendercritical} && \subrtab{politics} \\
        && \subrtab{thenewright} && \subrtab{worldnews} \\
        \midrule
        \multirow{3}{*}{\textit{cluster 3}} && \subrtab{chapotraphouse} && \subrtab{politics} \\
        && \subrtab{the\_donald} && \subrtab{askreddit} \\
        &&&& \subrtab{worldnews} \\
        \midrule
        \multirow{4}{*}{\textit{cluster 4}} && \subrtab{consumeproduct} && \subrtab{askreddit} \\
        && \subrtab{darkhumorandmemes} && \subrtab{politics} \\
        && \subrtab{darkjokecentral} && \subrtab{memes} \\
        && \subrtab{oandaexclusiveforum} && \\
        \midrule
        \multirow{2}{*}{\textit{cluster 5}} && \subrtab{wojak} && -- \\ 
        && \subrtab{imgoingtohellforthis2} && -- \\
        \bottomrule
	\end{tabular}
	\caption{Banned subreddits grouped by the cluster they belong to. For each cluster of banned subreddits, we report the corresponding most representative non-banned subreddits. Frequently occurring non-banned subreddits are used as baselines to compute the causal effect of the ban on the banned subreddits. Banned subreddits in cluster 5 are ignored as they are very small compared to the others.}
\label{tab:cluster_result}
\end{table*}
 
\subsection{Baseline group}
We obtain accurate causal estimates of the effects of The Great Ban by comparing the behavioral changes exhibited by the core users to possible changes exhibited during the same time by a set of reference Reddit users who were not affected by the ban. To reach this goal, we first identify a small set of related, but non-banned, subreddits to act as a baseline for the banned ones. To obtain this baseline, we first preprocessed every comment in \texttt{IN-BEFORE} for each banned subreddit by removing URLs, stopwords, and words less than four characters long. Subsequently, we performed stemming. For \subr{chapotraphouse} we executed these steps only on a random sample of 1M comments, due to the sheer number of comments in the subreddit. The preprocessed comments were then fed to BERTopic ~\citep{grootendorst2022bertopic} to extract up to ten relevant topics, each consisting of ten keywords. We then used the 100 descriptive keywords for each subreddit to project it with BERT~\citep{devlin-etal-2019-bert} into a 768-dimensional embedding space. This representation is suitable to highlight commonalities between the banned subreddits, such as via clustering, which we leverage to identify the small set of baseline subreddits. However, the high dimensionality of the embedding space is not suitable to directly perform clustering \citep{domingos2012few,zimek2018clustering}. Given the low number of data points -- the 15 banned subreddits -- we therefore first greatly reduce the dimensionality with PCA before clustering the banned subreddits with k-means \citep{ikotun2023k}. We experimented with multiple numbers $p$ of principal components and values for $k$, ultimately choosing $p=5$ and $k=5$ by optimizing the silhouette coefficient. Table \ref{tab:cluster_result} illustrates the clustering output. We now identify a set of reference non-banned subreddits for each cluster of banned subreddits, with the exception of \textit{cluster 5} that is formed by particularly small subreddits (i.e., \subr{imgoingtohellforthis2} and \subr{wojak}). For clusters one to four, we computed the cluster centroid and found the most representative keywords as those having the smallest Euclidean distance to the centroid. We then retrieved all comments posted during the pre-ban period that contained those keywords and identified the non-banned subreddits where such keywords occurred more frequently. These subreddits are related to the non-banned ones and are therefore suitable candidates for representing a baseline for our causal analyses. As reported in Table~\ref{tab:cluster_result}, \subr{politics}, \subr{askreddit}, and \subr{worldnews} are overall the three most representative non-banned subreddits for our clusters. 
As reported in Table~\ref{tab:dataset}, given the numerosity of the baseline subreddits, we selected a random stratified sample of 17K users, which is comparable to the number of core users in the focus group of banned subreddits.
From this set, we removed 45 users who posted at least one comment in one of the 15 banned subreddits, so as to avoid any overlap between the focus and baseline groups.

Selecting an appropriate baseline group presents inherent challenges, as different approaches come with trade-offs. While our method favors larger subreddits due to keyword-based selection, this choice ensures that the reference subreddits provide a stable and representative sample of general trends on the platform. An alternative approach, such as selecting smaller but thematically closer subreddits, could have introduced significant issues. For instance, \subr{donaldtrump} -- a seemingly relevant alternative for \subr{The\_Donald}, had minimal activity before the Great Ban and only surged afterward, making it both temporally inconsistent and dependent on the studied event. As a result, it would not serve as a valid counterfactual. Our selected baseline subreddits, while only approximating an ideal control group, strike a balance between ensuring thematic relevance and maintaining independence from the intervention, thus allowing for meaningful causal inference.

\subsection{Toxicity}
Several automated methods exist for assessing the toxicity of textual content. In this study, we compute toxicity scores using Detoxify~\cite{hanu2020detoxify}, an open-source deep learning classifier that provides reliable toxicity assessments. Detoxify achieves performance comparable to state-of-the-art solutions~\cite{cima2024great} and can be installed and executed locally. Due to its effectiveness and accessibility, it has been widely adopted in recent research on online toxicity~\cite{ejaz2024towards,kopf2024openassistant}. Among the various scores provided by Detoxify, we specifically use the \textit{toxicity} score, as it has been shown to exhibit the highest correlation with toxicity assessments from other established methods~\cite{cima2024great}.

 \clearpage{}\begin{table}[t]
    \footnotesize
    \setlength{\tabcolsep}{2pt}
    \centering
{\begin{tabular}{lrrrcrrrrrrrrrllrr}
        \toprule
	& \multicolumn{3}{c} {\textbf{aband. before} (ABA)} && \multicolumn{3}{c} {\textbf{remain. before} (BEF)} && \multicolumn{2}{c} {\textbf{remain. after} (AFT)}\\
	\cmidrule(lrr){2-4} \cmidrule(lrr){6-8} \cmidrule(lrr){10-11}
\textbf{subreddit} & \textit{users} & \textit{mean} & \textit{stdev} && \textit{users} & \textit{mean} & \textit{stdev} && \textit{mean} & \makecell[cb]{\textit{stdev}} \\
        \midrule
        \subrtab{chapotraphouse} & 886 & 0.163 & 0.123 && 8,319 & 0.145 & 0.084 && 0.138 & \makecell[cb]{0.094}  \\
        \subrtab{the\_donald} & 987 & 0.163 & 0.122 && 3,145 & 0.145 & 0.081 && 0.145 & \makecell[cb]{0.113} \\
        \subrtab{darkhumorandmemes} & 225 & 0.183 & 0.108 && 1,392 & 0.161 & 0.073 && 0.154 & \makecell[cb]{0.098} \\
        \subrtab{consumeproduct} & 444 & 0.190 & 0.090 && 1,275 & 0.158 & 0.045 && 0.160 & \makecell[cb]{0.099} \\
        \subrtab{gendercritical} & 333 & 0.184 & 0.119 && 706 & 0.167 & 0.071 && 0.141 & \makecell[cb]{0.116}  \\
        \subrtab{thenewright} & 151 & 0.177 & 0.092 && 575 & 0.153 & 0.076 && 0.145 & \makecell[cb]{0.098} &&  \\
        \subrtab{soyboys} & 162 & 0.200 & 0.074 && 432 & 0.177 & 0.072 && 0.163 & \makecell[cb]{0.087} \\
        \subrtab{shitneoconssay} & 171 & 0.190 & 0.084 && 384 & 0.165 & 0.071 && 0.160 & \makecell[cb]{0.098}\\
        \subrtab{debatealtright} & 148 & 0.171 & 0.084 && 328 & 0.153 & 0.070 && 0.155 & \makecell[cb]{0.115} \\
        \subrtab{darkjokecentral} & 38 & 0.221 & 0.154 && 270 & 0.155 & 0.070 && 0.144 & \makecell[cb]{0.092}\\
        \subrtab{wojak} & 70 & 0.192 & 0.068 && 170 & 0.160 & 0.065 && 0.143 & \makecell[cb]{0.044} \\
        \subrtab{hatecrimehoaxes} & 45 & 0.175 & 0.073 && 143 & 0.166 & 0.074 && 0.158 & \makecell[cb]{0.111}  \\
        \subrtab{ccj2} & 26 & 0.188 & 0.093 && 119 & 0.133 & 0.076 && 0.125 & \makecell[cb]{0.106} \\
        \subrtab{imgoingtohellforthis2} & 20 & 0.202 & 0.144 && 72 & 0.169 & 0.077 && 0.187 & \makecell[cb]{0.141} \\
        \subrtab{oandaexclusiveforum} & 4 & 0.129 & 0.107 && 55 & 0.188 & 0.086 && 0.182 & \makecell[cb]{0.092} \\
        \midrule
        focus group overall & 2,577 & 0.167 & 0.120 && 13,963 & 0.147 & 0.083 && 0.141 & \makecell[cb]{0.101}\\
        \midrule
        \subrtab{askreddit} & 274 & 0.146 & 0.093 && 15,576 & 0.121 & 0.066 && 0.120 & \makecell[cb]{0.072}  \\
        \subrtab{worldnews} & 143 & 0.146 & 0.083 && 9,907 & 0.125 & 0.067 && 0.124 & \makecell[cb]{0.072} \\
        \subrtab{politics} & 136 & 0.146 & 0.079 && 8,862 & 0.125 & 0.068 && 0.124 & \makecell[cb]{0.072} \\
        \midrule
        baseline group overall & 304 & 0.145 & 0.096 && 16,633 & 0.120 & 0.067 && 0.119 & \makecell[cb]{0.073} \\
	\bottomrule
        \multicolumn{13}{l} {*: $p < 0.1$; **: $p < 0.05$; ***: $p < 0.01$}
	\end{tabular}
	\caption{Subreddit-wise mean and standard deviation of toxicity scores for users who left Reddit after the ban (ABA) and for those who remained. For the latter, toxicity scores were computed both before (BEF) and after (AFT) the ban.
   }
   \label{tab:results}
}\end{table}
\clearpage{}

\clearpage{}\begin{table}[!t]
    \footnotesize
    \setlength{\tabcolsep}{2pt}
    \centering
{\begin{tabular}{l c c c c r c c c r c c c r c c}
        \toprule
        & & \multicolumn{2}{c}{\textbf{toxicity mean}} & & & & & & \multicolumn{6}{c}{\textbf{toxicity stdev}} \\
        \cmidrule{2-5}
        \cmidrule{8-15}
        \textbf{subreddit} & \multicolumn{2}{c}{ABA \textbf{\textit{vs}} BEF} & \multicolumn{2}{c}{BEF \textbf{\textit{vs}} AFT} & & & \multicolumn{4}{c}{ABA \textbf{\textit{vs}} BEF} & & & \multicolumn{3}{c}{BEF \textbf{\textit{vs}} AFT} \\
        \midrule 
        \subrtab{chapotraphouse} & $-0.018$ & *** & $-0.005$ & *** & & & & & $-0.047$ & *** & & & $+0.012$ & *** \\
        \subrtab{the\_donald} & $-0.018$ & *** & $-0.005$ & *** & & & & & $-0.046$ & *** & & & $+0.040$ & *** \\
        \subrtab{darkhumorandmemes} & $-0.022$ & *** & $-0.009$ & *** & & & & & $-0.047$ & *** & & & $+0.030$ & *** \\
        \subrtab{consumeproduct} & $-0.032$ & *** & $-0.009$ & *** & & & & & $-0.024$ & *** & & & $+0.029$ & *** \\
        \subrtab{gendercritical} & $-0.017$ & *** & $-0.019$ & *** & & & & & $-0.029$ & *** & & & $+0.023$ & *** \\
        \subrtab{thenewright} & $-0.024$ & *** & $-0.009$ & *** & & & & & $-0.018$ & *** & & & $+0.026$ & *** \\
        \subrtab{soyboys} & $-0.023$ & *** & $-0.011$ & *** & & & & & $-0.020$ & & & & $+0.012$ & *** \\
        \subrtab{shitneoconssay} & $-0.025$ & *** & $-0.007$ & *** & & & & & $-0.010$ & * & & & $+0.031$ & *** \\
        \subrtab{debatealtright} & $-0.018$ & ** & $-0.002$ & ** & & & & & $-0.017$ & *** & & & $+0.052$ & *** \\
        \subrtab{darkjokecentral} & $-0.066$ & *** & $-0.005$ & *** & & & & & $-0.103$ & * & & & $+0.024$ & *** \\
        \subrtab{wojak} & $-0.032$ & *** & $-0.011$ & *** & & & & & $-0.005$ & & & & $+0.044$ & *** \\
        \subrtab{hatecrimehoaxes} & $-0.009$ & & $-0.004$ & *** & & & & & $+0.020$ & & & & $+0.015$ & ** \\
        \subrtab{ccj2} & $-0.055$ & *** & $-0.008$ & *** & & & & & $-0.018$ & & & & $+0.033$ & *** \\
        \subrtab{imgoingtohellforthis2} & $-0.033$ & & $-0.004$ & ** & & & & & $-0.017$ & *** & & & $+0.065$ & *** \\
        \subrtab{oandaexclusiveforum} & $+0.059$ & & $-0.010$ & *** & & & & & $-0.024$ & & & & $+0.009$ & \\
        \midrule
        focus group overall & $-0.020$ & *** & $-0.006$ & *** & & & & & $-0.044$ & *** & & & $-0.077$ & *** \\
        \midrule
        \subrtab{askreddit} & $-0.025$ & *** & - & & & & & & $+0.033$ & *** & & & $+0.007$ & *** \\
        \subrtab{worldnews} & $-0.021$ & *** & - & & & & & & $+0.020$ & *** & & & $+0.006$ & *** \\
        \subrtab{politics} & $-0.021$ & *** & - & & & & & & $+0.016$ & *** & & & $+0.006$ & *** \\
        \midrule
        baseline group overall & $-0.025$ & *** & - & & & & & & $+0.037$ & *** & & & $+0.007$ & *** \\
        \bottomrule
        \multicolumn{9}{l}{* : $p < 0.1$; ** : $p < 0.05$; *** : $p < 0.01$}
    \end{tabular}
	\caption{Effect sizes of The Great Ban and their statistical significance. Mean toxicity results for BEF \textit{vs} AFT are obtained via Difference-in-Differences, while results for all other columns via direct subtraction. Statistical significance of the differences between mean values is obtained via t-tests, while that of differences in variability is obtained via F-tests on the variances.}
   \label{tab:results-DiD}
    }
\end{table}
\clearpage{}

\section{Analyses and Results}
\label{sec:results}
\subsection{RQ1: Effectiveness of The Great Ban}
\label{sec:results-rq1}
In this section, we analyze how effective The Great Ban was in reducing toxic behaviors.

\textbf{Abandoning users.} 
Table~\ref{tab:dataset} underlines a difference of 2,577 users between the \texttt{OUT-BEFORE} and \texttt{OUT-AFTER} datasets, meaning that (15.6\%) of the users became inactive following The Great Ban. The percentage of users who have abandoned the baseline subreddits is much smaller, at approximately 2\%, corresponding to 304 users. 
Since there is no sign of activity from these users during the seven months following the ban, we can presume that they not only abandoned Reddit but also likely migrated to other platforms.
This is the first clear effect of The Great Ban.
To provide more details, for each subreddit we compare the toxicity of abandoning users to the toxicity of those who remained on the platform.
As shown in Table~\ref{tab:results}, in 14 out of 15 subreddits, the abandoning users exhibited higher pre-ban toxicity than the others.
This means that toxic users were more inclined to abandon the platform after the ban, compared to their less toxic fellows.
In 12 subreddits, the difference in toxicity between users who abandoned and those who did not is statistically significant ($p < 0.05$) according to a t-test for unpaired data. Additionally, abandoning users exhibited significantly larger \textit{standard deviation}, as reported in Table~\ref{tab:results-DiD}, suggesting greater variability in toxicity.

\textbf{Checking model assumptions.} To estimate the causal effect of The Great Ban, we rely on Difference-in-Differences (DiD), which compares outcome changes over time between a treated and a control group. Here, we refer to the treated and control groups as the focus and baseline groups, respectively. This approach relies on the assumption that, without treatment, both groups would follow a common trend, allowing us to isolate the treatment effect. The following OLS regression model is used to estimate the effects:
\begin{equation}
\label{eq:did}
Y_{it} = \beta_0 + \beta_1 T_t + \beta_2 S_i + \beta_3 (T_t \cdot S_i) + \epsilon_{it}
\end{equation}
where \(T_t\) is a time indicator (\(1\) after the intervention, \(0\) before), \(S_i\) identifies the treatment group (\(1\) if treated, \(0\) otherwise), and the interaction term \(\beta_3\) captures the estimated effect. In practice, the DiD estimator for the effect can be computed as:
\begin{equation}
\delta = (y_{22} - y_{12}) - (y_{21} - y_{11})
\end{equation}
where \(y_{st}\) represents the outcome for group \(s\) at time \(t\). Additionally, $t = 1$ refers to the pre-ban period while $t = 2$ refers to the post-ban period, and $s = 1$ indicates the baseline group while $s = 2$ indicates the focus group. In addition to the standard OLS assumptions, the DiD model in Eq.~\eqref{eq:did} relies on the Stable Unit Treatment Value Assumption (SUTVA), ensuring that no spillover effects exist between treated and control groups, and on the parallel trends assumption, which states that, in the absence of intervention, both groups would have followed similar trends over time. We assume SUTVA to be valid based on our choice of baseline subreddits, which are completely non-overlapping it terms of users with the banned subreddits. Furthemore, we empirically verified the parallel toxicity trends between the focus and baseline groups by analyzing the monthly pre-intervention toxicity levels, which confirmed the parallel trends.

\textbf{Remaining users.} 
Besides leading a subset of users to abandon the platform, the ban may have also contributed to shifts in toxicity among the remaining users.
Table~\ref{tab:results}
presents, for each subreddit, toxicity scores for the set of remaining users, both before and after the ban.
It shows how, for the most part, the remaining users reduced their toxicity scores after the ban.
In particular, the scores aggregated by subreddit reveal that users from all 15 subreddits exhibited a slight decrease in their toxicity.
Table~\ref{tab:results-DiD} presents the size and statistical significance of the effects of the The Great Ban in terms of mean toxicity and its standard deviation, for all banned subreddits. Statistical significance of the differences between the means is obtained via t-tests, while that of dispersion is computed via F-tests on the variances. 
The results in Table~\ref{tab:results-DiD} indicate a consistent decrease in mean toxicity between abandoning and remaining users across all banned subreddits, with most differences being statistically significant ($p < 0.01$). This suggests that users who remained on Reddit exhibited lower toxicity levels than those who left after the ban.
Differences in standard deviation show mixed effects. When comparing abandoning and remaining users before the ban, we note that the latter showed less variability. This result is statistically significant ($p < 0.01$) in the majority of subreddits. Instead, when comparing remaining users before and after the ban, we find that the variability in toxicity increases after the intervention. This indicates that the ban has increased variability in the behavior of the remaining users, a phenomenon that we explore more thoroughly with subsequent analyses.
We also note that the effect sizes measured when comparing remaining users before and after the ban are notably smaller than those obtained comparing the abandoning users to remaining ones before the ban. In fact, after The Great Ban, there was an average reduction in overall toxicity of 4.08\% --- a relatively modest decrease.
To further validate our findings, we conducted the same analyses using the user-level frequency of toxic comments (i.e., those with toxicity scores $>0.5$) instead of the mean toxicity. We then compared these results with those reported in Tables~\ref{tab:results} and~\ref{tab:results-DiD}, finding that the results remained largely consistent, which reinforces the observed effects.

\begin{figure}[t]
    \centering
    \includegraphics[width=0.7\columnwidth]{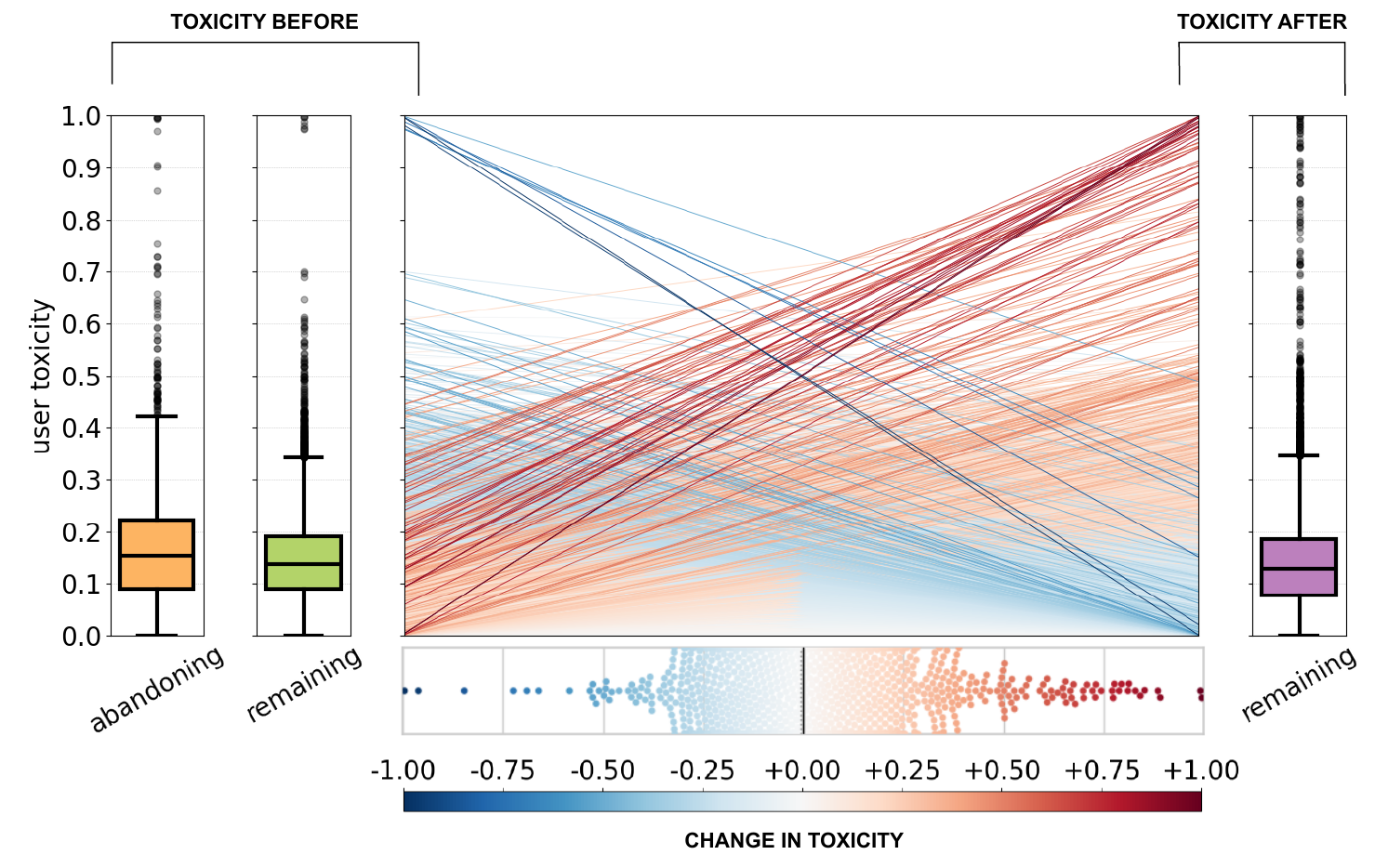}\caption{User-level toxicity changes after The Great Ban for each active user. The slope chart in the central panel highlights a majority of red-colored rising lines, indicating a substantial number of users who significantly increased their toxicity. The beeswarm plot in the bottom panel reinforces this observation, showing a higher concentration of users in the right red-colored tail of the distribution compared to the left blue-colored tail. The boxplots illustrate the marginal distributions for abandoning and remaining users, before and after the intervention.}
    \label{fig:slope-beeswarm-all}\end{figure}

\textbf{User-level effects.} 
Until now, we presented results at community-level, showing that after the ban, 15.6\% of core users abandoned the platform and the remaining ones reduced slightly their average toxicity by 4.08\%. To provide a more in-depth analysis, we also examine user-level effects by analyzing the changes in toxicity experienced by each of the 13,963 remaining users. The central panel of Figure~\ref{fig:slope-beeswarm-all} shows a slope chart representing user-level toxicity changes across all users, regardless of the subreddit in which they were active. A single line is associated with each user and line slopes represent the amount of increase or decrease in average toxicity. Rising lines are colored in different shades of red based on their slope, representing users who became more toxic after the ban.   Instead, the decreasing lines are in different shades of blue and represent users who reduced their toxicity. Figure~\ref{fig:slope-beeswarm-all} also features marginal boxplots displaying the toxicity distributions of remaining users before (left side of the slope chart) and after (right side) the ban. The leftmost boxplot displays the toxicity distribution for users who left the platform. Finally, the panel at the bottom shows the distribution of the changes in user-level toxicity in the form of a beeswarm plot. This is useful to identify outliers and study both tails of the distribution -- those associated with significant toxicity increases (red dots on the right-hand side of the beeswarm plot) and decreases (blue dots on the left-hand side).

The three boxplots in Figure~\ref{fig:slope-beeswarm-all} align with and support the general toxicity trends detailed in Table~\ref{tab:results}. Users who abandoned the platform have the largest average toxicity with respect to those who stayed. The group displaying the lowest average toxicity is the one formed by the remaining users after the ban.  The slope chart in Figure~\ref{fig:slope-beeswarm-all} shows that most lines have relatively small positive or negative slopes. This means that most users manifested little changes in toxicity, whether it is a change in positive or negative. At the same time, the slope chart also displays a notable number of steep lines, representing users who showed significant changes in toxicity. Specifically, the steep red lines outnumber the blue ones. This means that the majority of users who drastically changed their toxicity became more toxic. \rev{This result is consistent across multiple toxicity indicators -- including average toxicity, proportion of toxic comments, and absolute toxic comment volume -- and remains valid even when focusing only on users with stable and substantial activity before and after the ban.} This finding qualitatively illustrates an unintended consequence of The Great Ban: it led a significant minority of users to become resentful, resulting in markedly more toxic behavior. The points where the lines intersect the \textit{y} axis on the right side of the plot indicate the users' toxicity levels after the ban. This is also shown in the boxplot on the right side of the slope chart. As illustrated in Figure~\ref{fig:slope-beeswarm-all} and as suggested by the standard deviation values in Table~\ref{tab:results}, there is increased variability in user toxicity following the ban. What led to this increase in variability is the presence of the aforementioned resentful users. \rev{Additional correlation analyses further confirm that these users not only exhibited higher toxicity scores but also posted a higher raw number of toxic comments, reinforcing that the observed shifts reflect real behavioral changes rather than artifacts of reduced post-ban activity.}

Figure~\ref{fig:slope-beeswarm-all} presents the aggregated results across all subreddits. Nonetheless, the same plots can be used to investigate the effects of The Great Ban among single subreddits, as done in our earlier work~\cite{cima2024great}. This is useful for identifying common patterns and potential differences among the subreddits. We therefore conducted the analysis separately for users in each subreddit. The comparison across different subreddits reaffirmed earlier findings that most users who experienced significant toxicity changes after the ban increased their toxicity. However, this behavior was more pronounced in some subreddits and less evident in others. Overall, we observed that the findings for RQ1 in~\citep{cima2024great}, which utilized the median, are similar to those presented in this section when using the mean and a causal inference method such as DiD.

\multicomment{
\begin{figure}[t]
    \centering
    \includegraphics[width=\columnwidth]{img/the_donald_v2.pdf}
    \includegraphics[width=\columnwidth]{img/consumeproduct_v2.pdf}
    \caption{User-level toxicity changes for participants in \subr{the\_donald} (top) and \subr{consumeproduct} (bottom).}
    \label{fig:slope-beeswarm-donald-consume}\end{figure}

\begin{figure}[t]
    \centering
    \includegraphics[width=\columnwidth]{img/control-all_v2.pdf}\caption{User-level toxicity changes for participants in the control group.}
    \label{fig:slope-beeswarm-control-all}\end{figure}
}

\subsection{RQ2: Extreme user reactions to The Great Ban}
\label{sec:results-rq2}

Findings from RQ1 suggest that The Great Ban led to a small decrease in toxicity, except for a group of particularly resentful users who drastically increased their toxicity.
We now investigate the presence of these users across the subreddits.

\begin{figure}[t]
    \centering
    \includegraphics[width=0.7\columnwidth]{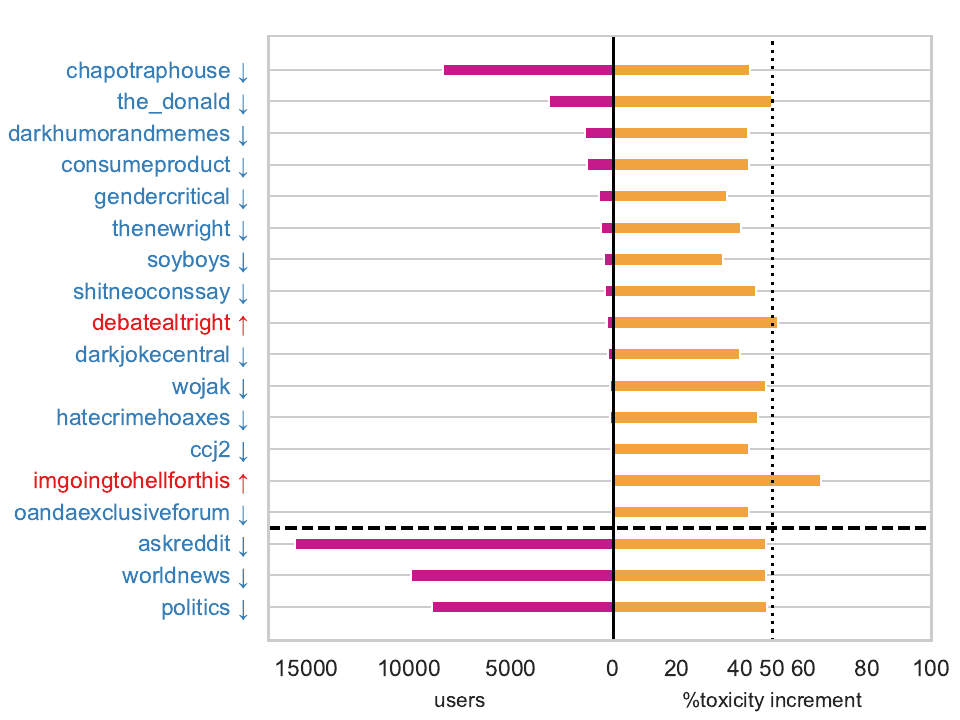}\caption{Subreddit-wise percentage toxicity increment (orange bars, right-hand side) and total number of users (purple bars, left-hand side), for the banned and baseline (non-banned) subreddits. Subreddits where the toxicity increases outweight the decreases are highlighted in red ($\uparrow$), while the others in cyan ($\downarrow$).}
    \label{fig:slope-beeswarm-control-all}\end{figure}

\begin{figure}[t]
    \centering
    \includegraphics[width=0.7\columnwidth]{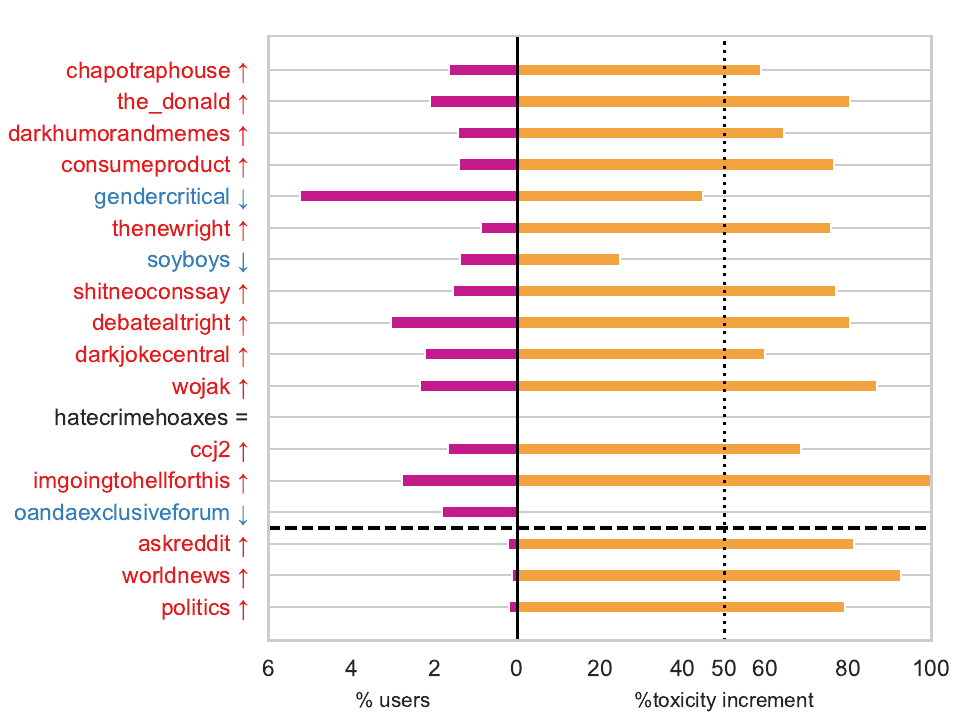}\caption{Subreddit-wise percentage toxicity increment of the outliers users (orange bars, right-hand side) and percentage of outlier users out of all subreddit users (purple bars, left-hand side), for the banned and baseline (non-banned) subreddits. Subreddits where the toxicity increases of the outlier users outweight the decreases are highlighted in red ($\uparrow$), while the others in cyan ($\downarrow$). The subreddit \subr{hatecrimehoaxes} is black-colored since it does not contain any outlier.}
    \label{fig:slope-beeswarm-control-out}\end{figure}

Let $t(i)_{\text{BEF}}, t(i)_{\text{AFT}}$ be the toxicity of the $i$-th user before and after the ban, and $\Delta t(i) = t(i)_{\text{AFT}} - t(i)_{\text{BEF}}$ be their change in toxicity.
The beeswarm plot of Figure~\ref{fig:slope-beeswarm-all} illustrates the distribution of $\Delta t(i)$ for all users.
We found that 5\% of all users have a $\Delta t(i) > 0.1$.
This finding is noteworthy considering that the mean toxicity pre-ban is 0.141, as reported in Table~\ref{tab:results}.
This means that 5\% of users increased their toxicity by more than 70\% after the ban.
To broaden the analysis, we compute the total change in toxicity within a subreddit with $N$ users as $\Delta t = \sum_{i = 1}^N \Delta t(i)$.
To evaluate separately the contributions of the two tails of the beeswarm plot, we only consider positive or negative $\Delta t(i)$. 
We now quantify and focus on the contribution of the right tail -- the one relative to increased toxicity -- as follows:
\[
\Delta t^+ = \sum\nolimits_{i = 1}^N \Delta t(i)\quad \text{with } \Delta t(i) > 0
\]
Figure~\ref{fig:slope-beeswarm-control-all} illustrates the percentage increment toxicity for the banned and baseline subreddits, along with the number of users in those subreddits.
The corresponding decrease in toxicity, which is relative to the left tail, can be easily computed as $\Delta t^- = 100 - \Delta t^+$.
By looking at the percentage toxicity increment for the banned subreddits, we can deduce that the contributions of the left and right tails are quite balanced across nearly all subreddits. For example, in the case of \subr{the\_donald} these contributions are perfectly balanced. This is also the same for the baseline subreddits.
For 13 out of 15 banned subreddits (cyan colored), the decrease in toxicity slightly exceeds the increase ($\downarrow$), which leads to the overall modest reduction in toxicity observed in RQ1.
However, we can observe the opposite trend in \subr{debatealtright} and \subr{imgoingtohellforthis2} (red colored $\uparrow$).
By repeating the same analysis by using the frequency of toxic comments instead of the mean toxicity, we obtained results closely matching those in Figure \ref{fig:slope-beeswarm-control-all}, showing robustness to the choice of a toxicity indicator.

\multicomment{
\begin{figure}[t]
    \centering
    \includegraphics[width=0.7\columnwidth]{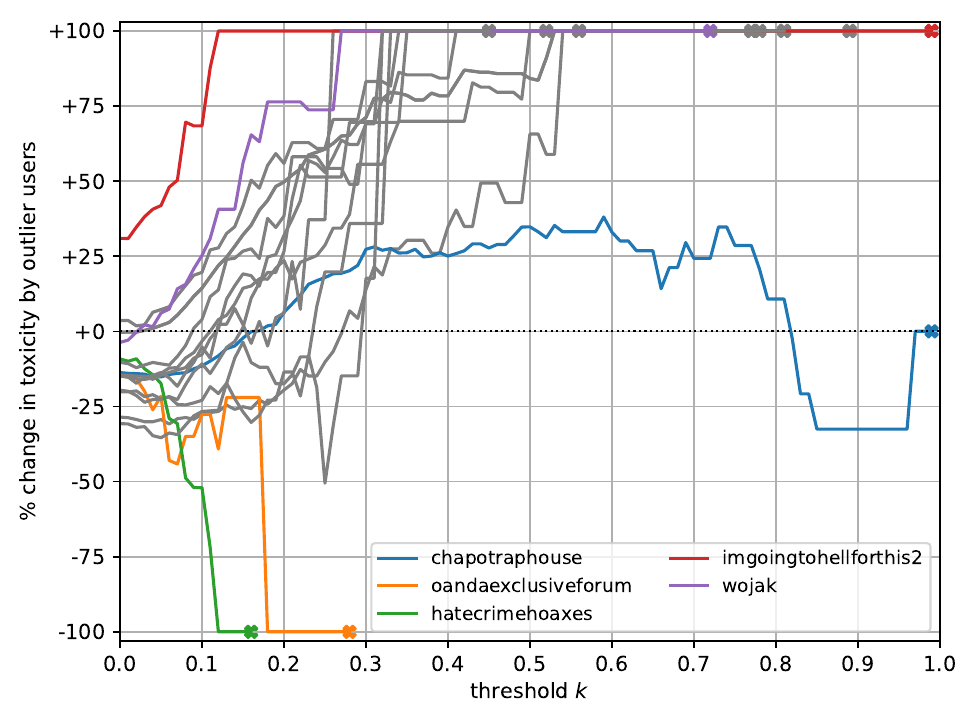}\caption{\hl{Contribution of the outlier users to the increase/decrease of toxicity in each subreddit. Outlier users are defined as those whose individual change in toxicity exceeds the threshold $k$. As shown, in 12 out of 15 subreddits the outlier users caused large toxicity increases.}}
    \label{fig:variation}\end{figure}
}

\begin{figure}[t]
    \centering
    \includegraphics[width=\columnwidth]{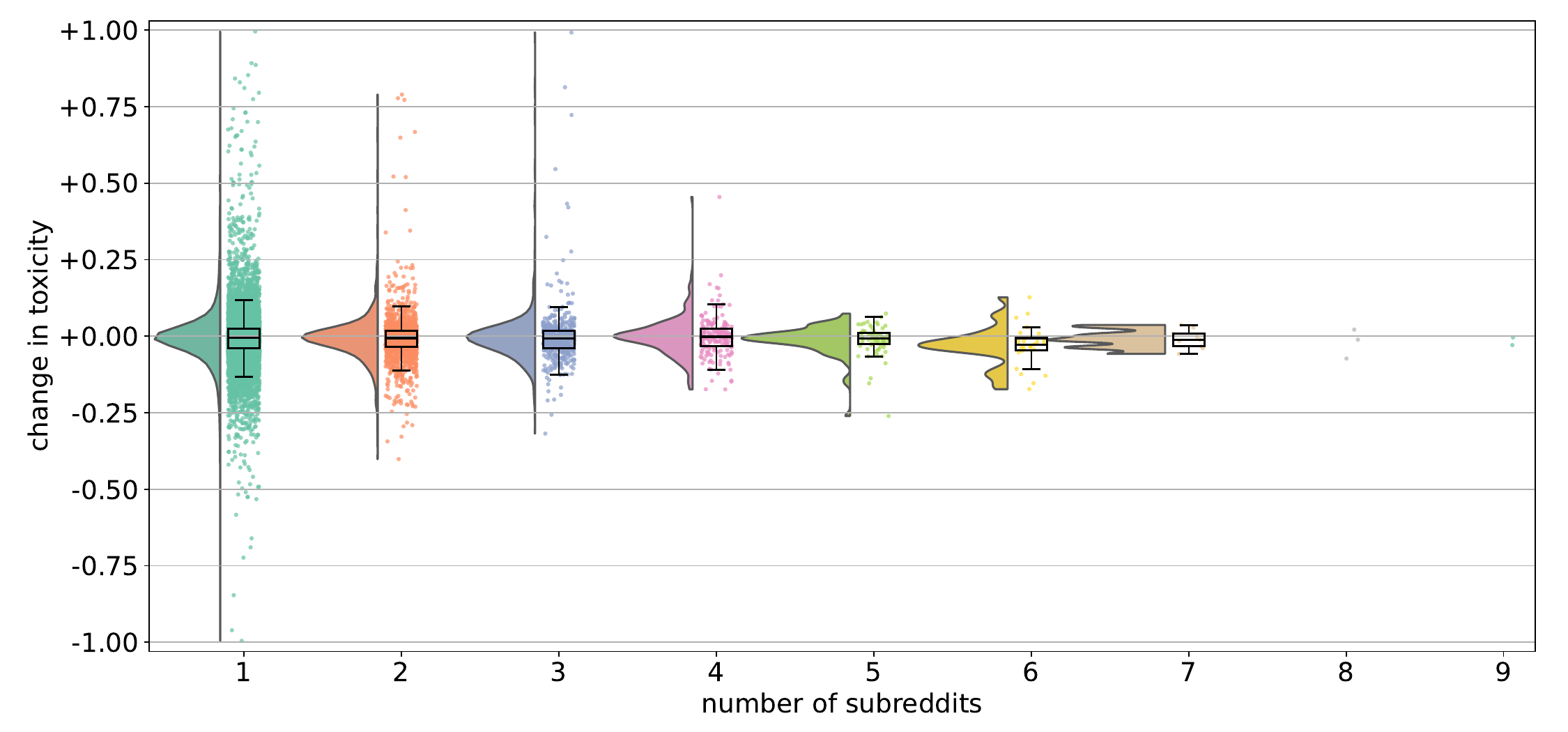}\caption{{Relationship between user participation in the 15 banned subreddits and change in toxicity. The vast majority of users only participated in one or two subreddits. Moreover, participation in more/less subreddits is unrelated to changes in toxicity.}} 
    \label{fig:raincloud}\end{figure}

To further assess the behavior of outlier users -- those who drastically increased or decreased their toxicity -- we recompute $\Delta t^+$ and $\Delta t^-$ by only considering those users whose $|\Delta t(i)| >0.25$. 
Figure~\ref{fig:slope-beeswarm-control-out} reiterates the previous analysis for outlier users. 
The proportion of outlier users (left-hand side of the figure) remains relatively small in all subreddits, with \subr{gendercritical} exhibiting the highest percentage with almost 6\% of users, while \subr{hatecrimehoaxes} does not have outliers at all.
The results are notably different than those of Figure~\ref{fig:slope-beeswarm-control-all}, as for 11 out of 15 banned subreddits (red colored $\uparrow$) the increase in toxicity (right-hand side of the figure) significantly outweighs the decrease. This does not apply to \subr{gendercritical}, \subr{soyboys}, and \subr{oandaexlusiveforms}, as for the first two subreddits, the increase in toxicity is less than 50\% and no increase is observed for the latter. For what concerns the baseline subreddits, the vast majority of outliers exhibited a marked toxicity increment. However, the number of outliers within the baseline subreddits is much lower with respect to the outliers in the banned subreddits, and ranges from 0.13\% to 0.22\%. In addition to these results, we also repeated the analysis for multiple values of $|\Delta t(i)|$, as described in~\cite{cima2024great}. We found that the obverved behavior is independent of the way in which outlier users are defined, which supports the robusteness of our finding. Overall, our results reveal that the phenomenon of outlier users who drastically increased their toxicity post-ban is peculiar of the banned subreddits, and can possibly be explained as a negative reaction of such users to the ban itself. 
Moreover, our findings also underscore that extremely toxic users are spread across almost all banned subreddits, which is a sign of a systemic reaction to the ban.

\subsection{RQ3: Relationship between toxicity, activity, and engagement}
\label{sec:results-rq3}
Our analyses in RQ2 revealed the widespread presence of extremely toxic users and quantified the extent of the issue across the subreddits. Here we move forward by providing additional information on their behavior in order to draw insights into their possible impact on the platform.

\textbf{Subreddits participation.} Initially, we explore the subreddit participation habits of these users as a function of their change in toxicity. This analysis aims to ascertain whether each of the most toxic users engaged in a few or many of the banned subreddits. Figure \ref{fig:raincloud} suggests that only a small number of users participated in multiple subreddits. Instead, the vast majority of users participated only in a few subreddits, typically ranging between one and three, consistently with the echo chamber theory~\citep{cinelli2021echo,valensise2023drivers}. Moreover, user toxicity changes do not depend on the number of subreddits in which the user participated. 

\begin{figure*}[t]
    \begin{minipage}[t]{0.49\columnwidth}\centering
        \includegraphics[width=\columnwidth]{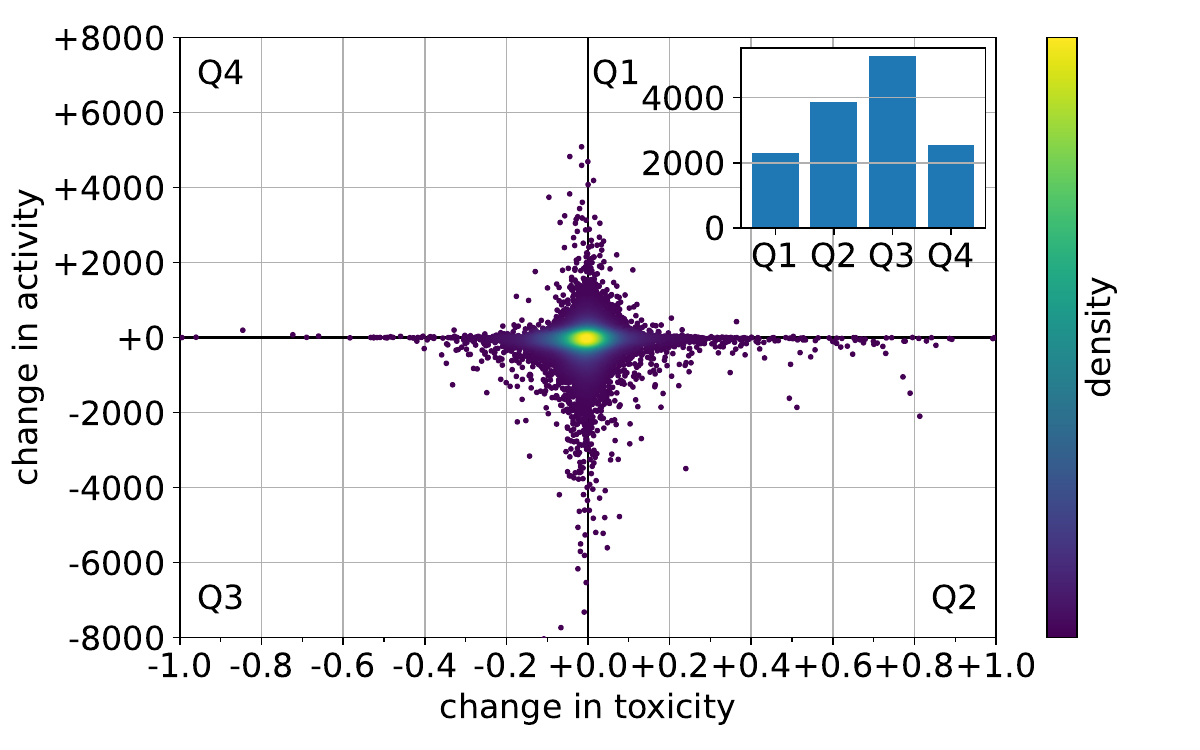}\end{minipage}
    \hspace{0.5pt}
    \begin{minipage}[t]{0.49\columnwidth}\centering
        \includegraphics[width=\columnwidth]{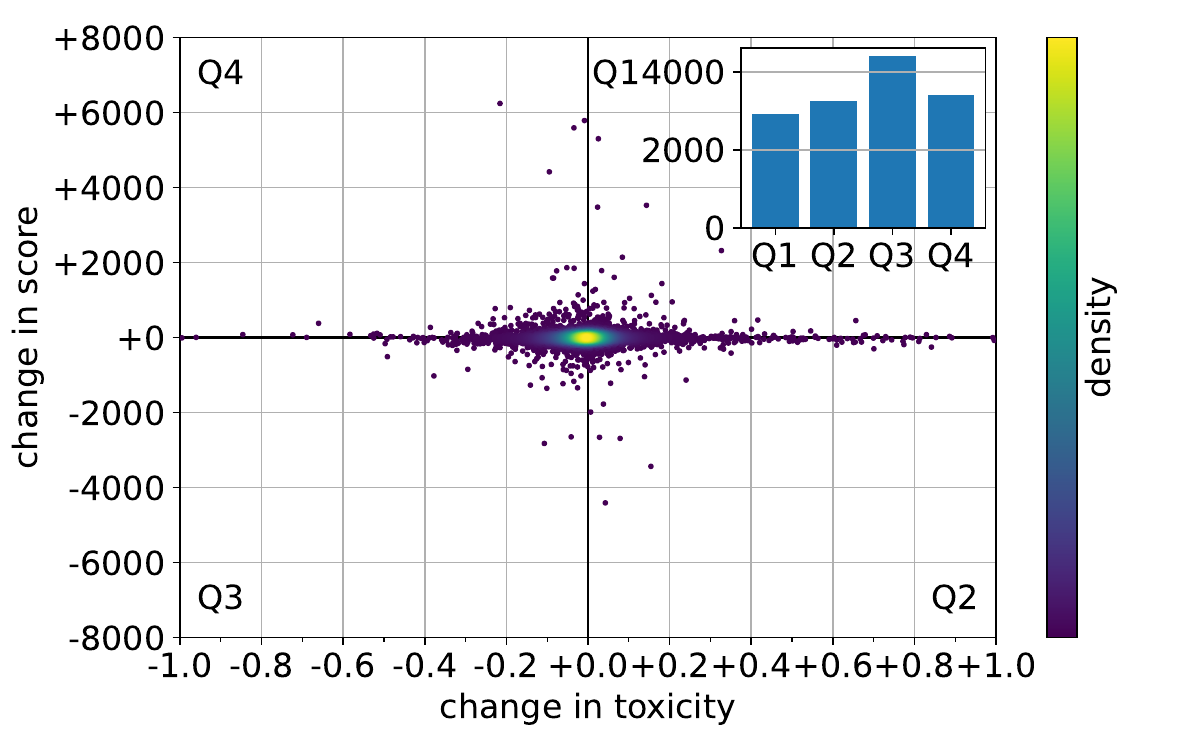}\end{minipage}\caption{Density scatterplots of the relationship between changes in toxicity and changes in activity (left-hand side) and score (right-hand side), for each user. The insets provide counts of the number of users in each quadrant of the scatterplots.
    }
    \label{fig:tox-act-eng}
\end{figure*}

\begin{figure*}[t]
    \begin{minipage}[t]{0.49\columnwidth}\centering
        \includegraphics[width=\columnwidth]{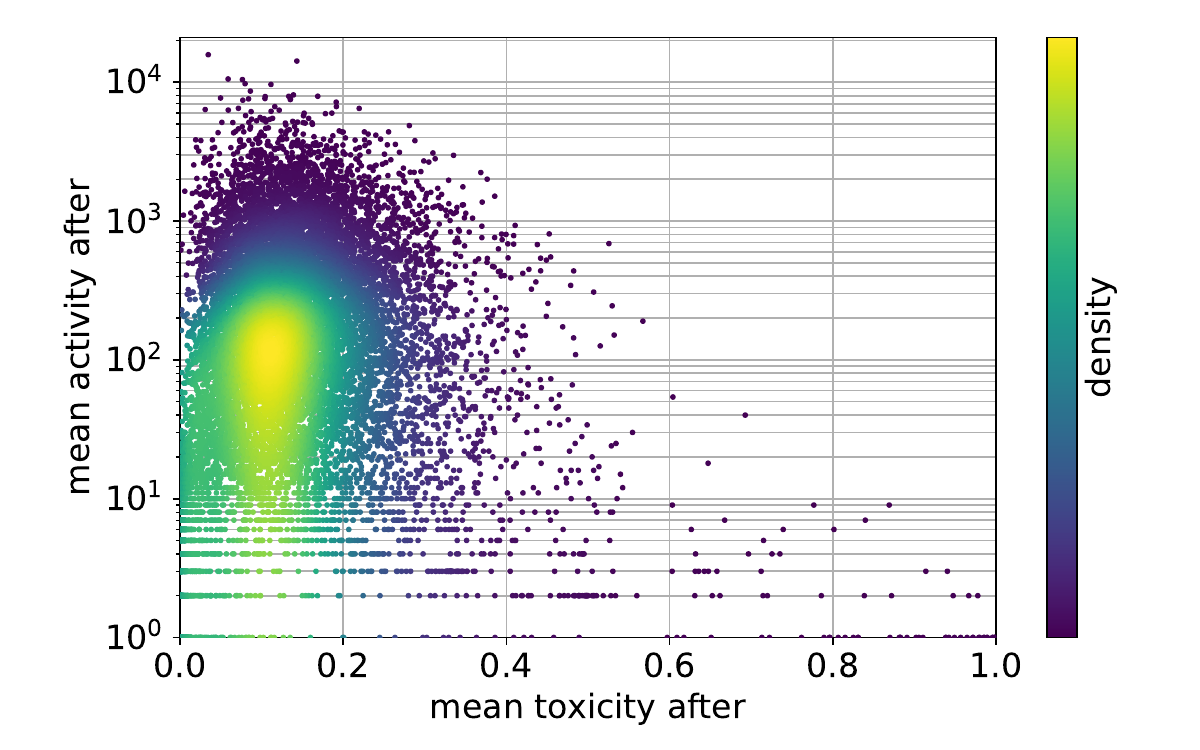}\end{minipage}
    \hspace{0.5pt}
    \begin{minipage}[t]{0.49\columnwidth}\centering
        \includegraphics[width=\columnwidth]{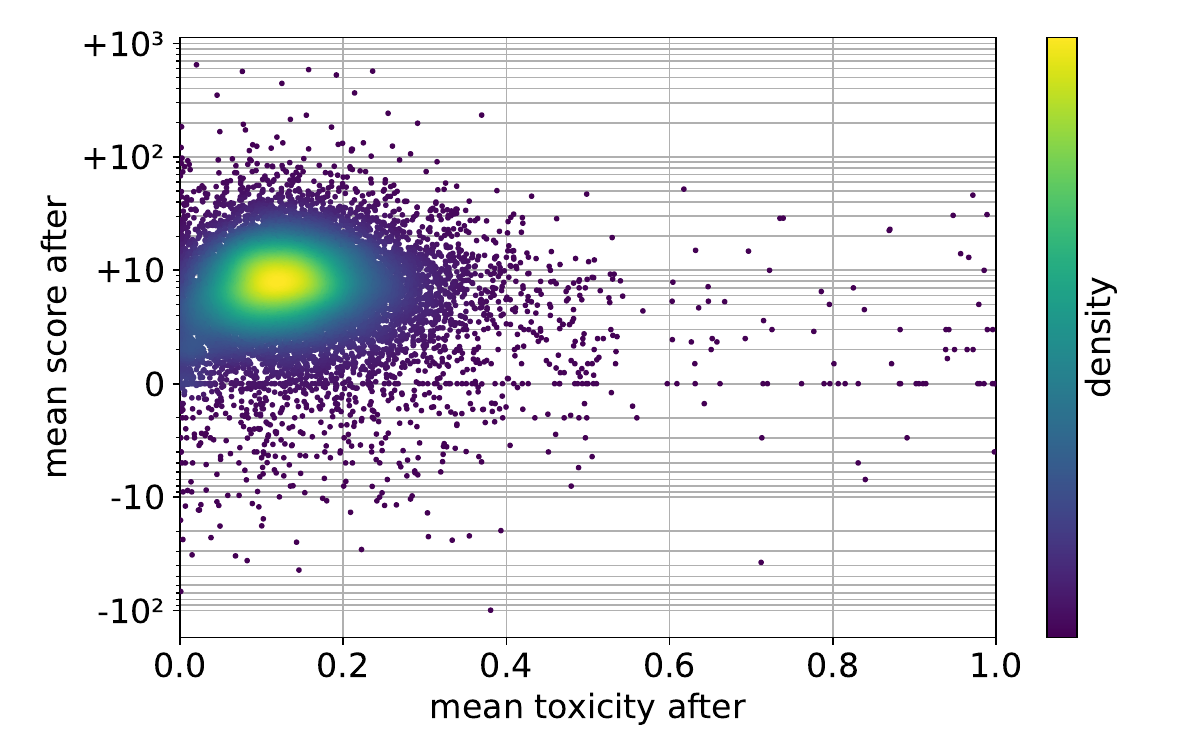}\end{minipage}\caption{Density scatterplots of the relationship between toxicity and activity (left-hand side) and score (right-hand side) after the ban, for each user. The score measures the difference between the total number of upvotes and downvotes received by a user and is a proxy for the engagement received. }
    \label{fig:tox-act-eng-aft}
\end{figure*}

\textbf{Activity.} Next, we look into the relationship between changes in toxicity and activity. We measure user activity as the number of posted comments. Conjointly studying user activity and toxicity is important since toxic users pose a meaningful threat only when high toxicity coincides with elevated activity, which holds the potential to cause a substantial impact on the platform. This joint analysis also enables the identification of those users who are both highly toxic and remarkably active, allowing to prioritize moderation interventions towards them~\citep{cresci2022personalized}. The left-most plot of Figure~\ref{fig:tox-act-eng} presents a density scatterplot of the changes in toxicity and activity for each considered user. The plot area is divided into four quadrants, corresponding to the combinations of increase/decrease in toxicity/activity. As shown, the bulk of the distribution lies close to the plot origin, meaning that the ban caused minor changes in both toxicity and activity. Moreover, those users who experienced marked changes in toxicity only had negligible changes in activity, and vice versa. The insets in Figure~\ref{fig:tox-act-eng} show the number of users laying in each quadrant of the scatterplots. The most frequent reaction to the ban was a reduction in both toxicity and activity (38\% users lay in Q3). This is also evident in the left-most plot of Figure~\ref{fig:tox-act-eng-aft}, which pictures the relationship between mean toxicity and user activity, after the ban. It shows how the vast majority of users who exhibit low levels of toxicity are moderately active. The opposite behavior is instead the least frequent (16\% users are in Q1). Overall this analysis suggests that the resentful users might have had a limited impact on the platform due to their low activity. 

\textbf{Engagement.} Finally, we examine the relationship between changes in toxicity and obtained engagement. We measure the engagement obtained by each comment as its Reddit score, computed as the difference between the upvotes and downvotes to that comment. Positive scores denote a positive engagement of the community with a comment, while negative scores generally denote disagreement. The engagement of a user is the mean engagement obtained by its comments. This joint analysis of toxicity and engagement is relevant as it allows an understanding of whether post-ban toxic comments are tolerated, and possibly even encouraged, on Reddit. The right-most plot of Figure~\ref{fig:tox-act-eng} shows the density scatterplot of this relationship while the right-most plot of Figure~\ref{fig:tox-act-eng-aft} shows the distribution of mean toxicity with respect to the mean engagement score after the ban.
Similarly to what occurred for activity in Figures~\ref{fig:tox-act-eng} and~\ref{fig:tox-act-eng-aft}, we found that the vast majority of users experienced modest changes in toxicity and engagement, and lower values of engagement scores correspond to lower levels of toxicity. Again, the most frequent reaction was a simultaneous reduction of both toxicity and obtained engagement (31\% users are in Q3) and the least frequent one was a simultaneous increase of both (21\% users are in Q1).

\begin{figure*}[t]
    \begin{minipage}[t]{0.49\columnwidth}\centering
        \includegraphics[width=\columnwidth]{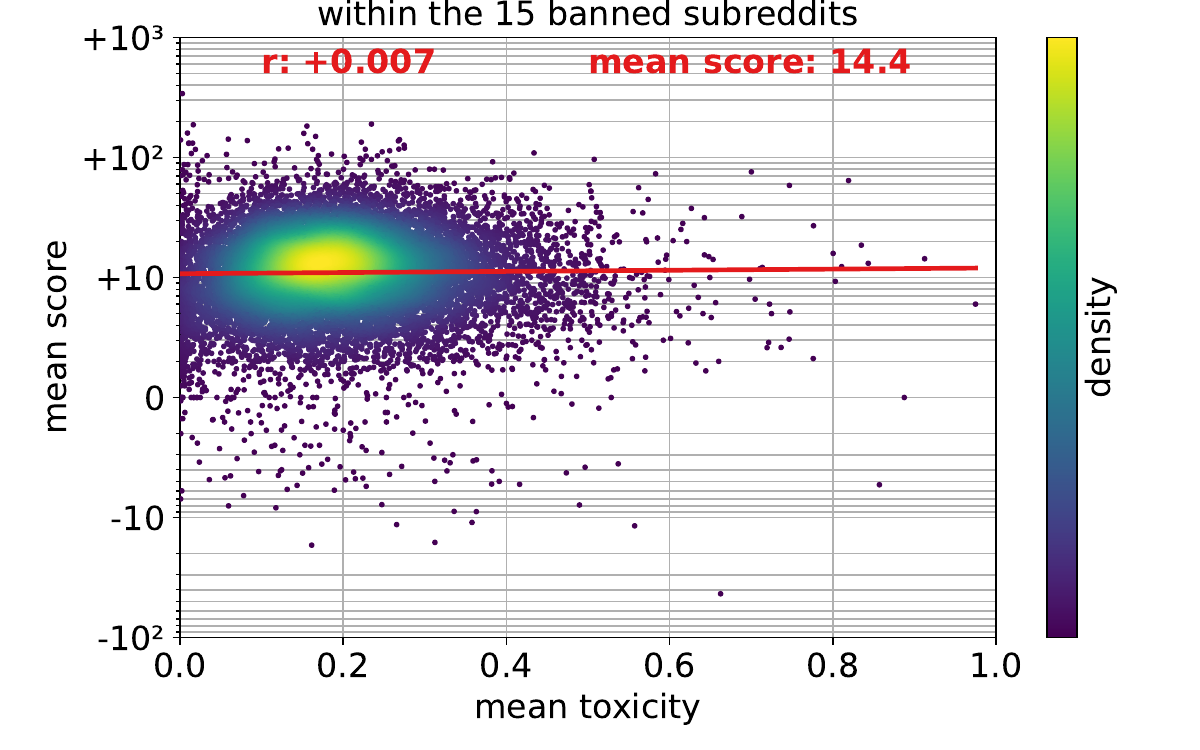}
    \end{minipage}
    \hspace{0.5pt}
    \begin{minipage}[t]{0.49\columnwidth}\centering
        \includegraphics[width=\columnwidth]{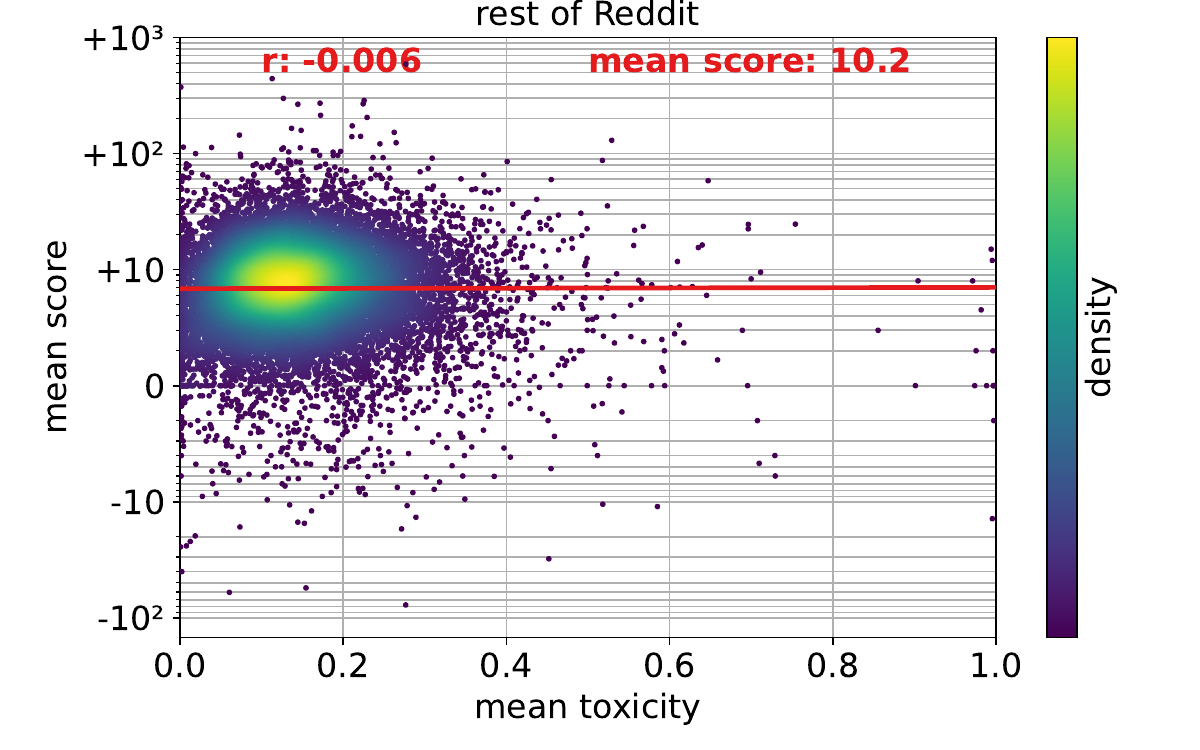}
    \end{minipage}\caption{Density scatterplots of the relationships between toxicity and score within the banned subreddits (left-hand side) and in the rest of Reddit (right-hand side). For each scatterplot, we also report the Pearson correlation $r$ between toxicity and score, and the mean score.}
    \label{fig:score}
\end{figure*}

Lastly, we analyze the relationship between toxicity and engagement before the ban, by investigating differences between the engagement obtained within \textit{vs} outside of the 15 banned subreddits. Figure~\ref{fig:score} provides the results of this comparison, which shows that toxic comments received slightly more engagement within the banned subreddits than in the rest of Reddit. This is evident from the higher concentration of toxic content in these communities, where the toxicity score distribution is more skewed toward large values. Additionally, comments in the banned subreddits received significantly higher mean scores than those elsewhere ($p < 0.01$), suggesting that engagement with toxic content was strong and that such content was not actively discouraged. We further examined the correlation between toxicity and user-assigned scores, finding very weak yet opposite trends: a positive Pearson correlation ($r = +0.007$) in banned subreddits and a negative one ($r = -0.006$) in the rest of Reddit. While not statistically significant, this pattern suggests that toxicity might have been tolerated or even slightly incentivized within banned communities, whereas it was mildly discouraged elsewhere. This result corroborates that of the right-most plot of Figure~\ref{fig:tox-act-eng} in that it reinforces the idea that toxicity is not encouraged on Reddit, except for a few toxic subreddits.

 \section{Discussion}
\label{sec:discussion} 
Our study provides insights into the consequences of The Great Ban, a crucial example of deplatforming that affected over 2,000 subreddits.
We highlight that 15.6\% of the users involved abandoned the platform, however those who remained reduced their average toxicity by 4.1\%.
At the same time, however, approximately 5\% of all users significantly escalated their toxicity.
The presence of these so-called resentful users was observed across many of the analyzed subreddits. 
However, only 16\% of users increased both their toxicity and activity levels, and just 21\% of users who became more toxic also received positive feedback from other users.
Our findings provide new insights regarding the impact of The Great Ban and, in general, on the heterogeneous and unintended negative reactions to moderation interventions~\citep{trujillo2023one}. Additionally, they shed light on several aspects regarding the design and implementation of effective moderation strategies, underlining the inherent challenges in regulating online platforms.

\subsection{Effectiveness of the moderation} 

In the current literature, the effectiveness of content moderation has typically been evaluated by examining the changes occurred after certain moderation interventions in terms of \textit{activity} and \textit{toxicity}~\citep{chandrasekharan2017you,jhaver2021evaluating}.
In this context, our study showed that a significant portion of toxic users left the platform, while those who remained exhibited a modest decrease in toxicity along with a notable decline in activity.
Apparently, these results suggest successful moderation. However, it is crucial to carefully examine these findings by considering potential unintended consequences.
For instance, the deplatforming of toxic users may move the issue somewhere else rather than solve it.
As highlighted by recent studies, moderated users might migrate to other platforms~\citep{horta2021platform}, possibly indicating that their toxic interactions may have been displaced rather than mitigated.
Moreover, migrated users may engage in much more toxic and aggressive behavior in the new platforms~\citep{horta2021platform}.
Finally, user abandonment and reduced activity after the ban could represent a significant issue for Reddit, as user engagement and interactions are the main sources of revenue for online platforms~\citep{trujillo2022make,tessa2024beyond}.
As a result, the presumed effectiveness of The Great Ban in reducing toxicity should be interpreted carefully as it can negatively impact not only the platform's revenue because of abandonment or reduced engagement, but also outside Reddit because of user migration.
Future efforts should focus on finding a balance between developing strategies that effectively reduce toxicity while minimizing user abandonment or decline in activity \citep{tessa2024beyond}.
In fact, effective moderation should foster healthier online communities while safeguarding platform safety and economic sustainability. Otherwise, platforms may be discouraged from implementing rigorous moderation measures.

\subsection{Divergent reactions to moderation} 
\rev{Our study showed that, despite a slight decrease in toxicity at the aggregate level, a significant minority of users exhibited substantial increases in their toxic behavior. This finding has important implications for evaluating and designing moderation interventions. First, it underscores the complexity of user responses to content moderation, an area that remains vastly underexplored and warrants further investigation~\citep{jhaver2023personalizing}. Additionally, it brings to light the necessity for \textit{personalization} in the context of content moderation~\citep{cresci2022personalized}. A broad intervention like The Great Ban -- which affected thousands of subreddits and tens of thousands of users -- may not adequately address the diverse motivations and behavioral patterns of those impacted. Here, this was demonstrated by the subset of users who reacted with markedly increased toxicity. Our analyses confirm that these escalations were not artifacts of sparse post-ban activity or indicator instability: users who became more toxic often did so both in terms of average toxicity and in the absolute number of toxic comments. This reinforces the interpretation that a genuine shift occurred in the behavior of a small segment of users. Understanding what drives such divergent reactions is crucial for developing more nuanced and effective moderation interventions. Future research and real-world implementations should therefore focus on user profiling, taking into account individual traits, behavioral history, and contextual factors to support more adaptive and personalized moderation strategies~\citep{cima2025contextualized,annamoradnejad2022multi}. In addition, the observed user migration -- combined with a sizable fraction of users displaying increased toxicity -- raises concerns about the potential \textit{radicalization} effects of moderation and its unintended role in reinforcing echo chambers~\citep{horta2021platform}. These findings suggest that some users may not respond positively to specific forms of intervention, and may instead engage in more extreme behaviors. Overall, our results emphasize the careful balance needed in content moderation: interventions should aim not only to reduce toxicity at the community level but also to anticipate and mitigate the risk of backlash among vulnerable user groups.}

\subsection{Impact of resentful users}
The finding that the vast majority of resentful users -- who exhibited increased toxicity following the ban -- did not concurrently escalate their activity offers a novel perspective on the little-studied interplay of toxicity and user activity. Contrary to the suspicion that highly toxic users might also be highly active, our results underscore a limited overlap between these two dimensions of user behavior. This suggests that a considerable portion of users expressing heightened toxicity post-ban did not translate their resentment into increased participation on the platform. In turn, this result partly mitigates the previous concerns about the divergent and adverse reactions to The Great Ban. Furthermore, the observation that users who intensified their toxic behaviors did not receive heightened positive social feedback from their peers resonates with some recent research on the drivers of online hateful and toxic speech. The latter posits that online toxicity may be driven by the pursuit of social approval rather than by a desire to harm~\citep{jiang2023social}. To this end, our results demonstrate that the Reddit community did not endorse or support the increased toxicity exhibited by these users. According to the aforementioned theories, toxic users who do not receive the expected levels of social approval should decrease their toxicity over time~\citep{jiang2023social}. However, we measured overall stationary toxicity trends over the course of the 7 months after the ban. Our results thus apparently challenge the social approval theory, in support of competing theories according to which at least a subset of toxic and hateful users seek amusement by harassing others, as in trolling behavior~\citep{chandrasekharan2022quarantined,uyheng2022language}. On the one hand, these results call for additional research to identify the types and characteristics of toxic users. On the other hand, they reinforce the need for nuanced and diversified moderation strategies that are capable of addressing the complexity and multiplicity of online user behavior~\citep{cresci2022personalized}.

\subsection{Legal and regulatory implications}
Our findings reveal the moderate effectiveness of The Great Ban. While the intervention led to a mild overall reduction in toxicity, it also resulted in heterogeneous effects: a subset of users became significantly more toxic, while a considerable portion of previously active users abandoned the platform altogether. These mixed outcomes align with broader findings in the literature, suggesting that moderation interventions do not always produce the intended or necessary effects~\citep{bail2018exposure,pennycook2020implied,horta2021platform,trujillo2022make}. These results raise critical questions about the effectiveness and accountability of content moderation policies, particularly in the context of evolving regulatory frameworks in both the US and Europe.

In the US, our results connect to ongoing debates surrounding Section 230 of the Communications Decency Act~\citep{section230}, which provides platforms with broad discretion in how they moderate content. Notably, Section 230 does not require platforms to moderate content at all, nor does it hold them accountable for the effectiveness of their moderation efforts. The fact that The Great Ban produced only marginal improvements in reducing toxicity, while simultaneously exacerbating toxicity in a subset of users, underscores a key issue in the current legal landscape: platforms are not obligated to implement moderation strategies that are evidence-based or demonstrably effective. These findings contribute to recent discussions on potential reforms to Section 230, which aim to increase platform responsibility for harmful content~\citep{armijo2021reasonableness,citron2023fix}. Such reforms could incentivize the development of more strategic, proactive, and adaptive moderation frameworks~\citep{habib2022proactive,tessa2024beyond}, shifting away from reactive mass-deplatforming toward targeted interventions that minimize unintended consequences~\citep{cresci2022personalized,cima2025contextualized}.

On the other hand, the European Union’s Digital Services Act (DSA) takes a markedly different approach by imposing stricter transparency and accountability requirements on content moderation~\citep{eu2020DSA}. Our findings -- particularly the observation that a fraction of users became more toxic and resentful following the ban -- relate to key DSA provisions that mandate platforms to assess the risks of their moderation policies, including unintended side effects, before enforcement. While this requirement represents an important step toward more responsible platform governance, significant uncertainties remain regarding: \textit{(i)} how platforms currently evaluate the risks associated with their moderation policies, \textit{(ii)} whether they possess the capacity to effectively measure or mitigate these risks, and \textit{(iii)} the extent to which opaque moderation decisions impact users and broader platform dynamics. These challenges call for renewed multidisciplinary research efforts to better assess the unintended consequences of content moderation and to develop evidence-based mitigation strategies that align with evolving regulatory expectations~\citep{tessa2024beyond,cima2025contextualized}.

In conclusion, our study highlights how the current regulatory landscape of content moderation varies significantly across regions. The US model prioritizes platform autonomy, allowing platforms to moderate at their own discretion without clear accountability for their outcomes. In contrast, EU regulations emphasize transparency, risk assessment, and harm mitigation, pushing platforms to justify and evaluate their moderation decisions more rigorously~\citep{trujillo2023dsa,kaushal2024automated,tessa2025improving,shahi2025year}. As global challenges in platform governance continue to evolve, platforms may increasingly need to tailor their moderation strategies to comply with diverging regional requirements -- a shift that could shape the future of online content regulation worldwide.

\subsection{Limitations and Future Works}
In this study, we make use of a large dataset consisting of comments posted by 34K Reddit users during seven months before and after The Great Ban. This means that our findings may be limited to Reddit and the specific circumstances of this intervention. Moreover, online platforms are dynamic and characterized by frequent changes in user behavior, community norms, and platform policies. Our study takes into account a fixed timeframe -- while relatively extended, it constrains the generalizability of our findings to other periods and evolving trends. \rev{In addition, our evaluation of toxicity relies on Detoxify, a machine-learning classifier whose output can be sensitive to the volume of comments per user. To guard against this potential artifact, we conducted robustness checks using alternative toxicity measures (e.g., proportion and absolute count of toxic comments) and restricted analyses to users with stable, substantial activity. These additional experiments confirmed that our main conclusions -- namely, a modest overall toxicity decline alongside a small subset of users who escalated their toxic behavior -- hold under different definitions and activity constraints.} The observational nature of our work also constitutes a limitation, as it reduces our ability to account for external events that might have independently influenced user behavior. For instance, major socio-political events such as the George Floyd protests in mid-2020, which sparked widespread discussions on racism and online hate speech~\cite{reny2021opinion}, may have contributed to shifts in both user behavior and moderation enforcement. Likewise, the COVID-19 pandemic, which heavily influenced online interactions and engagement patterns during 2020, could represent another confounding factor. While our Difference-in-Differences (DiD) design and inclusion of a baseline group mitigate such concerns to some extent, we acknowledge that these external factors may have shaped the observed behavioral changes. Finally, data on user demographics, motivations, and contextual factors is missing from our dataset. Such information could have offered a more granular interpretation of our results. Future work should explore the connections between user characteristics and moderation outcomes, supporting the development of more targeted, personalized interventions~\citep{cresci2022personalized,tessa2024beyond,cima2025contextualized}. \rev{Moreover, our methodology for selecting the baseline subreddits introduces an additional limitation. Our keyword-based approach tends to favor larger subreddits, which may introduce a size-related bias compared to the typically smaller banned communities. While this choice ensures greater statistical stability and robustness of observed trends, it may reduce thematic comparability. Alternative options—such as smaller, thematically closer subreddits—were considered but often violate the Stable Unit Treatment Value Assumption (SUTVA) due to overlap with the treated groups, thereby compromising causal validity. Future research could further explore this trade-off by identifying alternative reference groups based on user demographics, interaction patterns, or pre-ban engagement similarity, to better balance topical relevance and statistical independence.}

\subsection{Ethical considerations} 
This research furthers our understanding of the effects of content moderation by highlighting the complexities of user responses to moderation efforts. 
These new insights can contribute to the development of more robust and targeted interventions for reducing online toxicity while minimizing unintended side effects.
Our study also discusses the ethical trade-off between the common good and minority harm. Specifically, the dilemma faced by moderators who must choose whether to enforce actions that, while benefitting the broader community, may inadvertently cause harm to a minority of users by increasing their resentment.
 \section{Conclusions}
\label{sec:conclusions}
The Great Ban was a large-scale deplatforming intervention aimed at shutting down toxic communities on Reddit. To assess its effectiveness, we analyzed 53M comments made over 14 months by almost 34K users involved in the ban.
Our findings show that 15.6\% of the affected users left Reddit following the ban, while those who remained reduced their toxicity by an average of 4.1\%.
Despite this modest reduction in overall toxicity, 5\% of users increased their toxicity by more than 70\% compared to their levels before the ban.
While the presence of resentful users was pervasive in the analyzed subreddits, their impact in terms of activity and obtained engagement was limited. We found that only 16\% of users increased both their toxicity and activity and only 21\% also received positive engagement from the community.

Our results provide new and more detailed insights into the effectiveness of The Great Ban, as well as its unintended consequences.
Our findings can be beneficial to platform administrators and policymakers in the development of new moderation interventions and regulatory policies.
For example, future works could investigate the relationship between user characteristics and the effects of a moderation intervention.
This could lead to the development of targeted or personalized approaches that can be more effective in limiting the negative effects of moderation actions, such as the ones found in our study.
Furthermore, an encouraging direction for future research is the implementation of predictive models that forecast the effects of moderation interventions. Such models would be beneficial to content moderation, as they would allow moderators to plan and estimate the effectiveness of certain moderation strategies ahead of their deployment.

\section*{Acknowledgments}
This work was supported by the European Union -- Next Generation EU, Mission 4 Component 1, for project PIANO (CUP B53D23013290006); by the ERC project DEDUCE under grant \#101113826; by the PNRR-M4C2 (PE00000013) “FAIR-Future Artificial Intelligence Research" - Spoke 1 "Human-centered AI", funded under Next Generation EU; and by the Italian Ministry of Education and Research (MUR) in the framework of the FoReLab projects (Departments of Excellence).

\section*{Declaration of generative AI and AI-assisted technologies in the writing process}
During the preparation of this work the authors used \texttt{ChatGPT 4o} to proofread the article prior to its submission. After using this tool, the authors reviewed and edited the content as needed and take full responsibility for the content of the published article.

\bibliographystyle{elsarticle-harv} 
\bibliography{references}

\end{document}